\documentstyle[12pt,psfig]{article}
\begin{document}
\thispagestyle{empty}
\setcounter{page}{1}

\begin{center}
{\large \bf Polaron Variational Methods In The Particle
Representation Of Field Theory :  I. General Formalism}

\vspace{1cm}
R.~Rosenfelder $^{\star}$ and A.~W.~Schreiber $^{\star \> \>
\dagger}$

\vspace{1cm}
$^{\star}$ Paul Scherrer Institute, CH-5232 Villigen PSI, Switzerland

$^{\dagger}$ TRIUMF, 4004 Wesbrook Mall, Vancouver, B.C.,
Canada V6T 2A3

\end{center}

\vspace{1.5cm}
\begin{abstract}
\noindent
We apply nonperturbative
variational techniques to a relativistic scalar field theory in
which
heavy bosons (``nucleons'') interact with light scalar mesons
via a Yukawa coupling.
Integrating out the meson field and neglecting the nucleon vacuum
polarization one obtains an effective
action in terms of the heavy particle coordinates
which is nonlocal in the proper time.
As in Feynman's polaron
approach we approximate this action by a retarded quadratic action
whose parameters are to be determined variationally on the pole of
the two-point function.
Several ans\"atze for the retardation function are studied and for
the most general case we derive a system of coupled variational
equations. An approximate analytic solution displays the instability
of the system for coupling constants beyond a critical value.

\end{abstract}

\vspace{1.5cm}


\newpage

\section{Introduction}

\noindent
Variational methods have a long history and are still widely used in
physics to obtain approximate non-perturbative solutions.
For a very wide class of problems specified by a given set of
equations it is indeed always possible
to construct a variational principle which will give an estimate of
the quantity of interest correct to first order if the quantities
appearing
in the variational principle are known to zeroth order \cite{GRS}.
In quantum mechanics the best-known variational principle is the
Rayleigh-Ritz
variational principle for the energy which is applied extensively in
molecular, atomic and nuclear physics.

In contrast, the applications of variational principles in quantum
field theory are rather limited (for a review see Ref. \cite{Hay}).
Within the Hamiltonian formalism several studies exist
(~see, e.g.,\cite{DSSS,DiDa}~). The best known covariant
example is also a Rayleigh-Ritz variational principle
 which has been formulated in the functional Schr\"odinger
representation \cite{CJT}. It leads to the Hartree (-Fock)
approximation
when a gaussian wave functional is used. Unfortunately the latter is
the only trial functional which can be used for practical purposes,
 which drastically restricts the power of the variational principle.
In addition, in quantum field theory it is not the energy of the
ground state
(vacuum) one is interested in but the energy (mass) of excitations.
Already in ordinary
quantum mechanics this is much harder to obtain. The need for
renormalization
and the infinitely many degrees of freedom add to the ``difficulties
in applying the variational principle to quantum field theory''
so that Feynman expressed a rather pessimistic view on a workshop
devoted to that topic \cite{Fey2}.

It is remarkable that the variational principle works very well
in a nonrelativistic field-theoretical problem, the polaron (for
reviews see \cite{GeLo,BoPl,MCM,RoFe}), but only after the
infinitely many degrees of freedom for the
phonons are integrated out exactly. This gives rise to a non-local
effective action which Feynman approximated variationally by a
retarded quadratic action \cite{Fey1}. Recent exact Monte-Carlo
calculations \cite{AlRo} have again demonstrated that the Feynman
polaron approximation is the best analytical approximation which
works for small as well as large coupling constants. Taking the
known strong-coupling
expansions as a yardstick the ground-state energy deviates less than
$ 2.2 \% $ and the effective mass (which determines the lowest
excitations) less than $ 12 \%$ from the exact values.
This success can be attributed to
the reduction in the number of variables and the explicit allowance
of retardation in the quadratic trial action. Feynman used a specific
parametrization for the retardation function but the most general
form gives only a very small improvement in the ground state energy
\cite{AGL,Sai}.

Although the Feynman variational principle (or Jensen's inequality
in mathematical language) has sometimes been used in field theory
(see e.g. \cite{DiPe}),
it was never used in the context which made it so successful in the
polaron problem: namely, approximating a nonlocal action expressed
in terms of particle coordinates by a retarded quadratic one. We will
 do so in the present work which is the first in a planned series
about variational
approximations employing the {\it particle representation}
of field theory. The concept of using particle trajectories as
dynamical variables in a relativistic quantum theory is an old one:
it dates back to the 1937 paper by Fock \cite{Fock} who
investigated the role of proper time
in relativistic equations. In the early 50's Nambu \cite{Nambu},
Feynman \cite{FeyQED} and Schwinger \cite{Schwi} made much use of it,
but canonical (``second'') quantization later took over and dominated,
 in particular in the text books (an exception is, of course,
Ref. \cite{FeHi} ). Only a few works \cite {Frad,Den,HaSi} have
employed this approach in the following years.
The renewed interest in the particle representation
(see also \cite{KaKt,McRe}) is due to superstring-inspired techniques
for efficient calculation of one-loop diagrams which have been shown
to be connected to the (``first'')-quantized form of field theory
\cite{Str}.

For the moment we want to restrict our discussion to {\it scalar}
field theories. This
avoids the complications of spin in a path integral,
for which there is extensive discussion in the literature
(see, for example, \cite{BDH,FrGi}).
Also, having in mind applications in few-body physics, we take
the simplest field theory where a light scalar particle (the ``pion'') has a
Yukawa coupling to a heavy scalar (the ``nucleon''). This is the
Wick-Cutkosky model \cite{Wick,Cut} which usually is considered
as a simple model for relativistic bound-state problems treated in the ladder
approximation to the Bethe-Salpeter equation
(see e.g. Ref. \cite{ItZu}, Chapter 10-2).
Recently it also has become a popular playing
field for light-cone techniques \cite{Saw,Hil,WiHi}.

\begin{figure}
\unitlength1mm
\begin{picture}(100,70)
\put(0,0){\makebox(100,80){\psfig{figure=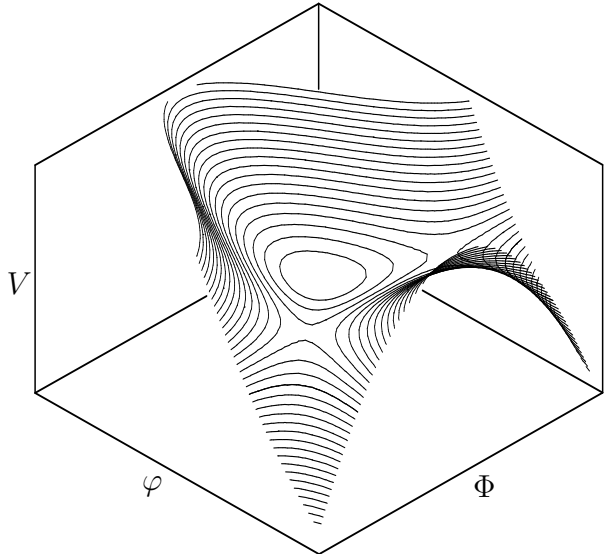,height=300mm,width=500mm}}}
\put(112,3){$\Phi$}
\put(68,4){$\varphi$}
\put(50,30){$V$}
\end{picture}
\caption{Contour plot of the classical ``potential''
of Eq.~(\protect\ref{classical potential}).}
\label{fig: contourplot}
\end{figure}

To be specific we consider the following Lagrangian in euclidean
space time
\begin{equation}
{\cal L} = \frac{1}{2} \left ( \partial_{\mu} \Phi \right )^2 +
\frac{1}{2}
M_0^2 \Phi^2 + \frac{1}{2} \left ( \partial_{\mu} \varphi \right )^2 +
\frac{1}{2} m^2 \varphi^2 - g \Phi^2 \varphi
\end{equation}
where $M_0$ is the bare mass of the heavy particle
(which we shall call, for brevity, the ``nucleon''), $m$ is the mass
of the
light particle (the ``meson'') and $g$ is the (dimensionfull) coupling
constant of the Yukawa interaction between the two particles. It is
well known \cite{Bay} that such a coupling
is equivalent to a $\Phi^3$ theory and therefore the ground state of
the theory is unstable. This is best seen in Fig.~\ref{fig: contourplot}
which shows a contour plot of the classical ``potential''
\begin{eqnarray}
V^{(0)}(\Phi,\varphi) &=& \frac{1}{2} M_0^2 \Phi^2 \> + \>
\frac{1}{2} m^2 \varphi^2 - g \Phi^2 \> \varphi \nonumber \\
&=&
\frac{1}{2} M_0^2 \Phi^2 \> - \frac{g^2}{2 m^2} \Phi^4
+ \> \frac{1}{2} m^2 \left ( \varphi - \frac{g}{m^2} \Phi^2 \right )^2
 \> .
\label{classical potential}
\end{eqnarray}
The superscript zero reminds us that this is the potential in
zeroth order in an expansion in powers of $\hbar$.
 One-loop quantum corrections modify the behaviour shown in
Fig.~\ref{fig: contourplot} somewhat, but no qualitative change occurs.

\begin{figure}
\unitlength1mm
\begin{picture}(100,80)
\put(0,0){\makebox(100,80){\psfig{figure=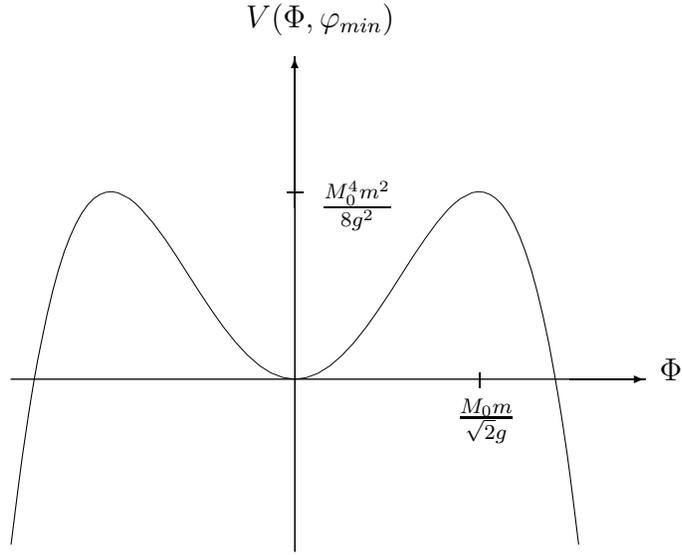,height=200mm,width=500mm}}}
\put(95,55){${M_0^4 m^2 \over 8 g^2}$}
\put(85,80){$V(\Phi,\varphi_{min})$}
\put(140,33){$\Phi$}
\put(113,27){${M_0 m \over \sqrt{2} g}$}
\put(128,33.15){\vector(1,0){10}}
\put(91.47,56){\vector(0,1){20}}
\put(116,32){\line(0,1){2}}
\put(90.5,58){\line(1,0){2}}
\end{picture}
\caption{Cut of the classical potential along the line
(\protect\ref{smallphi min}).}
\label{fig: double well}
\end{figure}

Clearly, the minimum at $ \Phi = \varphi = 0$ is only a {\it local}
minimum.
For positive $\varphi$ and nonzero $\Phi$ the ``potential'' decreases
indefinitely. Therefore  the `ground state' sitting near
$ \Phi = \varphi = 0$
is only {\it metastable}, at least in a classical  description.
 From semiclassical descriptions of tunneling \cite{Col,CaCol}
we expect
the lifetime to depend
on the minimum height and thickness of the barrier
for a given coupling constant.

Differentiating Eq. (\ref{classical potential}) with respect to
$\varphi$ we
obtain
\begin{equation}
\varphi_{\rm min} = \frac{g}{m^2} \Phi^2
\label{smallphi min}
\end{equation}
and the ``potential'' along this path
\begin{equation}
V^{(0)}(\Phi,\varphi_{\rm min}) = \frac{1}{2} M_0^2 \Phi^2 \> -
\frac{g^2}{2 m^2} \Phi^4
\label{inverted double well}
\end{equation}
is an {\it inverted} double well as shown in Fig.~\ref{fig: double well}.

We are, however, not genuinely interested in this instability of
the `ground
state' in the Wick-Cutkosky model.  Rather, we want to use it as a
field
theoretical  toy model for the dressing of {\it physical} nucleons
by mesons.
The arguments showing the instability of the `ground state' for a
scalar
``nucleon''  do not apply for the case where the nucleons have spin
\cite{Bay}. In other words, the instability is an unwanted side
effect of the simplified
model considered here and we shall ignore it whenever possible.
Operationally, we can  do this as long as we restrict the parameters
of the model such that
the width of the ground state is small compared to its mass. From the
 above arguments it is clear that this corresponds to sufficiently
small couplings.
Indeed it will turn out, quite reasonably, that the variational
equations we shall derive cease to have real solutions once the
coupling becomes too large;
i.e. the formalism itself tells us in which region it remains
applicable.

We will study the dressing of
a single ``nucleon'' in the quenched approximation, i.e. neglecting
pair creation of heavy particles which should be a good approximation
in low-energy processes. In this approximation it is possible to
integrate
out the mesons exactly and to obtain an effective non-local action
which is a covariant functional of the particle four-coordinates with
the proper time as parameter. This effective action bears a surprising
similarity to the polaron action so that we could even call the
dressed particle a ``relativistic polaron''.
We then perform a variational calculation with a quadratic trial
action in complete analogy to the polaron
case, except that we use a covariant description and have to
renormalize the mass of the heavy particle.
Recently Simonov and Tjon
\cite{SiTj,NTS} have also studied the Wick-Cutkosky model in the
quenched approximation and in the particle representation. However,
their aim was to solve the relativistic bound state problem beyond
the ladder approximation and they
neglected all self-energy and vertex corrections. Consequently there
is no need for renormalization and no sign of the instability in
their work.

This paper is organized as follows:
In Section \ref{sec: particle repr} and \ref{sec: var on 2point pole}
we respectively derive the effective
action in the particle representation of the Wick-Cutkosky model and
perform the variational approximation \`a la Feynman.  The latter
is done at the pole of the two-point function.
 In Section \ref{sec: var ansaetze} we discuss different variational
ans\"atze for the
retardation function and we set up the coupled system of equations
which arises when no assumptions are made about the form of the
retardation function. We study a simple approximate solution of these
variational equations which displays the instability of the ground
state. The main results of this work are
summarized in the last Section whereas some technical details are
relegated to the Appendix.

\section{Effective Action in the Particle Representation}
\label{sec: particle repr}

\noindent
We begin with the generating functional for the Green functions of
the theory,
\begin{equation}
Z\> [J,j] = \int {\cal D} \Phi  \> {\cal D} \varphi \>
\exp \left ( - S[\Phi,\varphi]
+ (J,\Phi) + (j,\varphi) \> \right )\;\;\;.
\label{generating}
\end{equation}
Here
\begin{equation}
S[\Phi,\varphi] = \int d^4x \> {\cal L}(\Phi(x),\varphi(x))
\end{equation}
denotes the action and we use
\begin{equation}
(J,\Phi) \equiv \int d^4x \> J(x) \Phi(x) \>\>\>\> {\rm etc.}
\end{equation}
as a convenient abbreviation for the source terms.

Our aim will be to integrate out the mesonic degrees of freedom in
order to get an effective action for the heavy particles. Indeed,
as the meson field
$ \varphi$ appears at most quadratically in the path integral one
could do so immediately, using
\begin{equation}
\int {\cal D} \varphi \exp \left [ - \frac{1}{2} (\varphi, D \>
\varphi) + (j,\varphi) \> \right ]
= \frac{{\rm const}}{({\rm det} \> D)^{1/2} } \>
\exp \left [ \> \frac{1}{2} \> (j,D^{-1} j) \> \right ] \> .
\label{gaussian path integral}
\end{equation}
Considering for simplicity the case $j = 0$, we'd obtain
\begin{equation}
\int {\cal D} \varphi \> \exp \left [ - \frac{1}{2}
(\varphi,(- \Box + m^2)\varphi) + g (\phi^2,\varphi) \> \right ] =
\frac{{\rm const}}{({\rm det}\> D_m)^{1/2}} \> \exp \left [
\frac{g^2}{2} \> (\Phi^2, \> D_m^{-1}, \Phi^2 ) \right ]\;,
\label{outint smallphi}
\end{equation}
where
\begin{equation}
 D_m \equiv - \Box + m^2
\label{inverse meson propagator}
\end{equation}
is the inverse meson propagator.
In Eq. (\ref{outint smallphi})  the prefactor arising from the
gaussian integration is independent of the field $ \Phi$ and
the sources and can be absorbed
in the (irrelevant) normalization factor for the path integral.
Therefore the effective action for the heavy field would be given
by
\begin{equation}
S_{\rm eff}[\Phi] = \frac{1}{2} ( \Phi,(-\Box + M_0^2) \Phi ) -
\frac{g^2}{2} (\Phi^2,\frac{1}{-\Box + m^2} \Phi^2) \> .
\end{equation}
This is a nonlocal $\Phi^4$-theory whose interaction term has the
wrong
sign, i.e. this action is not bounded from below. This leads to the
vacuum instability discussed above for the classical limit.
To solve the model completely
one now would still have to perform a functional integral over the
heavy field $\Phi$. Due to the non-gaussian nature of the resulting
path integral this is impossible to do analytically and one has
to resort to approximative methods.

Given that we want to apply a variational approach, it turns out
(as we shall see later) that it is actually advantageous to
first integrate out the heavy field before doing the same for
the light field. Although this sounds paradoxical in view of the
stated aim, we will reintroduce
the heavy particle {\it coordinate} at a later stage. Applying
Eq. (\ref{gaussian path integral}), we obtain
\begin{equation}
\int {\cal D} \Phi \> \exp \left [ - \frac{1}{2}
(\Phi, (-\Box + M_0^2 - 2g \varphi) \Phi) + (J,\Phi) \right]
= \frac{\rm const}{ \left [\> {\rm det}(-\Box + M_0^2 - 2g \varphi)
\> \right]^{1/2}} \> \exp (- I \>[\Phi,J] \> )
\end{equation}
with
\begin{equation}
I \> [\Phi,J] =  - \frac{1}{2} \> \left ( J, \> \frac{1}
{-\Box+M_0^2 - 2g \varphi} \> J \right ) \> .
\end{equation}
In contrast to Eq. (\ref{outint smallphi}) the prefactor now
explicitly depends on the meson
field $ \> \varphi\> $ over which have to finally integrate. As
the determinant is a highly nonlinear and nonlocal object
this makes an analytical
evaluation impossible. However, it is well known that the prefactor
describes pair production which is greatly suppressed
if the mass of these particles is large:
\begin{eqnarray}
\frac{{\rm det}(-\Box + M_0^2 - 2g \varphi)}{\rm const} &=&
\frac{ {\rm det}(-\Box+M_0^2 - 2g \varphi)}{{\rm det}(-\Box
+ M_0^2)} \nonumber \\
&=& {\rm det}
\left ( 1 - 2 g \frac{1}{-\Box + M_0^2} \varphi \right ) \> \>
\buildrel M_0 \to \infty \over \longrightarrow \> \> \> 1 \> .
\label{quenched approximation}
\end{eqnarray}
In the following we will adopt this ``quenched approximation''
and concentrate on the {\it two-point}
function for one nucleon with an arbitrary number of mesons.
For this object we
then have the following generating functional
\begin{eqnarray}
Z'\> [j,x] &\equiv& \frac{\delta^2 Z \> [J,j] }{\delta J(x) \>
\delta J(0)} \Biggr |_{J=0} \nonumber \\
&=& \int {\cal D} \varphi \> < x \> | \frac{1}
{-\Box + M_0^2 - 2g \varphi} |\> y = 0  >
\> \exp \left [ -\frac{1}{2} (\varphi, D_m \varphi ) + (j,\varphi)
\right ] \> .
\label{generating' field}
\end{eqnarray}
This obviously describes the propagation of a ``nucleon'' in the
presence of an external field $ \> \> g \> \varphi(x)\> $ over which
one has to integrate functionally with a given weighting function.
To perform this integration we use a trick due to
Schwinger and exponentiate the nucleon propagator
\begin{equation}
\frac{1}{ \hat p^2 + M_0^2 - 2g \varphi(x)} = \frac{1}{2}
\int_0^{\infty} d\beta \>
\exp \left [ - \frac{1}{2} \beta \> (\hat p^2 + M_0^2 -
2g \varphi(x) ) \> \right ]
\label{euclid proper time}
\end{equation}
where $\hat p_{\mu} = \partial_{\mu}/i $ is the four-momentum
operator.
The integration variable $\beta$ usually is called ``fifth
parameter'' or ``proper time''
(Refs. \cite{Fock,Nambu,FeyQED,Schwi}). Actually Eq.
(\ref{euclid proper time}) only holds if the corresponding
operator is positive definite which, in general, is {\it not} the
case since the meson field $\varphi(x)$ can take any values when
integrated over functionally. This means that the meson fluctuations
can become so large that the nucleon locally becomes massless or
even tachyonic. The correct way to exponentiate therefore would be
\begin{equation}
\frac{1}{ \hat p^2 + M_0^2 - 2g \varphi(x) -i \epsilon} =
\frac{i}{2} \int_0^{\infty} d T \>
\exp \left [- \frac{i}{2} T \> (\hat p^2 + M_0^2 - 2g \varphi(x)
- i \epsilon ) \> \right ]\> ,
\label{Mink proper time}
\end{equation}
i.e. to introduce Minkowski proper time instead of the euclidean
one as in
Eq. (\ref{euclid proper time}).  We recall from the Introduction
(see Eqs. (\ref{smallphi min}, \ref{inverted double well}) ) that
large meson fields can carry one over the
barrier and induce the instability of the ground state.
Since we want to disregard this instability
as much as possible and since numerical calculations
are much
easier in euclidean proper time we will nevertheless use Eq.
(\ref{euclid proper time}) in the following. However, we
should expect a breakdown
of this description for coupling constants large enough to induce
fluctuations over the barrier.

Even with the proper time representation (\ref{euclid proper time})
for the nucleon propagator we cannot perform the $\varphi$
integration
since the operator $\hat p^2$ does not commute with the external
potential $g \varphi(x)$. However, formally
\begin{equation}
U(x,\beta;0,0) =
 < x \> | \exp\left [ - \beta \> (\frac{\hat p^2}{2} -
g \varphi(x) ) \right ] | \> y = 0 >
\end{equation}
is the matrix element of the euclidean time evolution operator of a
{\it non} - relativistic particle of unit mass \footnote{A different
 value should not change {\it physical} observables since it only
corresponds to a different parametrization of the
particle path. It can be shown that such a  `reparametrization'
invariance holds in our variational approximation. The present
choice is called the `proper-time gauge' \cite{BDH}. } in the
potential $g \varphi(x)$.
Therefore we can express it as a path integral over the
{\it coordinate}
$x(\tau)$ of the particle beginning at $x(0) = 0$ and ending at
$x(\beta) = x$ \cite{FeHi,Schul}
\begin{equation}
U(x,\beta;0,0) = \int_{x(0)=0}^{x(\beta)=x} {\cal D}x(\tau) \>
\exp \left (
- \int_0^{\beta} d \tau \> \left [ \> \frac{1}{2} \dot x^2 -
g \varphi(x(\tau) ) \> \right ] \right ) \> .
\label{coordinate path integral}
\end{equation}
As all quantities in the path integral
(\ref{coordinate path integral}) are c-numbers the gaussian
$\varphi$-integral
\begin{equation}
\int {\cal D} \varphi \> \exp \left [ - \frac{1}{2} (\varphi, D_m
\> \varphi) \> + \> ( h,\varphi) \> \right ] \> ; \>\>\>\>\>
h(y) = \> j(y) \> + \> g \> \int_0^{\beta} d\tau \>
\delta(\> y - x(\tau)\>)
\end{equation}
can now be performed with the
help of Eq. (\ref {gaussian path integral}). The result is
\begin{equation}
Z'\> [j,x] = {\rm const} \int_0^{\infty} d \beta \>
\exp \left (- \frac{\beta}{2} M_0^2 \right )
\> \int_{x(0)=0}^{x(\beta)=x} {\cal D}x(\tau) \>
\exp ( - S_{\rm eff}\> [ x(\tau),j]\>  \> ) \;\;\;,
\label{generating' coordinate}
\end{equation}
where the effective action is given by
\begin{equation}
S_{\rm eff}\> [ x(\tau),j]\> = \int_0^{\beta} d \tau  \>
\frac{1}{2} \dot x^2 \> - \> \frac{1}{2} ( h, \> D_m^{-1}\> h )\> .
\end{equation}
It is convenient to write it in the form
\begin{equation}
S_{\rm eff}\> [x(\tau),j] = S_0\> [x(\tau)] \> + \>
S_1\> [x(\tau)] \> +\> S_2\> [x(\tau),j] \> + \> S_3\> [j]
\label{eff action}
\end{equation}
with
\begin{eqnarray}
S_0\> [x(\tau)] &=& \int_0^{\beta} d \tau  \> \frac{1}{2} \dot x^2
\label{eff action kin}\\
S_1\> [x(\tau)] &=&
- \frac{g^2}{2} \int_0^{\beta} d \tau_1 \> \int_0^{\beta}
d \tau_2 \> < x(\tau_1) |  \> D_m^{-1} \> | x(\tau_2) >
\label{eff action pot} \\
S_2\> [x(\tau),j] &=& - g \int d^4 y \> j(y) \> \int_0^{\beta}
d \tau \> < y | \> D_m^{-1} \> | x(\tau) >
\label{eff action source}\\
S_3\> [j] &=&  - \frac{1}{2} \int d^4 y_1 \> d^4 y_2 \> \> j(y_1) \>
< y_1 | \> D_m^{-1} \> | y_2 > \> j(y_2) \> .
\label{eff action disconnected}
\end{eqnarray}
Note that the last term $S_3\>[j]$ in the action does not depend on
the trajectory $x(\tau)$ of the nucleon and therefore the external
meson lines which are generated by differentiating with respect
to the meson source $j$
are not attached to the nucleon line. Thus the generating functional
for {\it connected} Green functions $G_{2,n}$ simply is
\begin{equation}
Z'_{\rm conn}\> [j,x] = Z'\> [j,x] \> \Bigr|_{S_3 = 0} \> .
\label{generating' connected}
\end{equation}
Compared to the usual procedure via a Legendre transform this simple
identification is just one of many advantages of field theory
in the ``particle representation''. Another one is the big reduction
in degrees of freedom: although in Eq. (\ref{generating' coordinate})
one still has to do a functional integration, it is over 4 functions
of one variable (the proper time), whereas the previous field
theoretical path integral (\ref{outint smallphi}) is over one
function of 4 variables (namely the space-time coordinates). It is
for this reason that one might
expect a variational approach based on particle coordinates to be
superior to the one based on field variables, given that in both
cases only quadratic trial actions can be used in practical
calculations.

Eqs. (\ref{eff action kin}, \ref{eff action pot}) are the
relativistic generalization of the retarded polaron action
which Feynman \cite{Fey1}
derived when integrating out the phonons from the polaron Hamiltonian.
The meson propagator may be written as
\begin{equation}
< x | \> D_m^{-1} \> | y > = \int \frac{d^4 q}{(2 \pi)^4} \>
\frac{e^{\> i q\cdot(x-y)}}{q^2 + m^2}\;\;\;,
\label{meson propagator}
\end{equation}
and so Eq. (\ref{eff action pot}) becomes
\begin{equation}
S_1[x(\tau)] =
- \frac{g^2}{2} \int_0^{\beta} d \tau_1 \> \int_0^{\beta} d \tau_2 \>
\int \frac{d^4 q}{(2 \pi)^4} \> \frac{1}{q^2 + m^2} \>
e^{ \> i q \cdot ( \>
x(\tau_1) - x(\tau_2) \> ) }  \> .
\label{eff action pot'}
\end{equation}
Comparing with the polaron action \cite{AlRo}
\begin{equation}
S_1^{\rm polaron}[x(\tau)]  = -\frac{\alpha}{2 \sqrt{2}}
\int_{0}^{\beta} d\tau_1 \int_{0}^{\beta} d\tau_2 \> \int
\frac{d^3 q}{2 \pi^2}
 \> \frac{e^{-|\tau_1 - \tau_2|}}{{\bf q}^2} \> e^{ \> i {\bf q}
\cdot ( \> {\bf x}(\tau_1)-{\bf x}(\tau_2) \> )}
\label{polaron action}
\end{equation}
one observes a striking similarity. This is even more pronounced
when we perform the $q_0$-integration in Eq. (\ref{eff action pot'})
which gives
\begin{equation}
S_1[x(\tau)] =
- \frac{g^2}{16 \pi} \int_0^{\beta} d \tau_1 \> \int_0^{\beta}
d \tau_2 \> \int \frac{d^3 q}{2 \pi^2} \>
\frac{e^{-\omega_q|x_0(\tau_1) - x_0(\tau_2)|}}{\omega_q} \>
e^{ \> i {\bf q} \cdot ( \>
{\bf x}(\tau_1) - {\bf x}(\tau_2) \> ) }
\label{eff action pot''}
\end{equation}
with $\omega_q = ({\bf q}^2 + m^2)^{1/2} $. However, there are also
some differences which should be noted :
\begin{description}
\item[~~(i)] All coordinates and momenta in $S_1$ in
Eq. (\ref{eff action pot'}), as opposed to $S_1^{\rm polaron}$,
 are four-dimensional and therefore
Lorentz invariance is explicit.
\item[~(ii)] A massive meson propagator enters into the effective
action of the Wick-Cutkosky model instead
of the Coulomb propagator in the polaron problem.
\item[(iii)] The explicitly Lorentz invariant expression for
$S_1$ ( Eq. (\ref{eff action pot'}) ) does not contain a retardation
factor in the proper time, whereas the
polaron effective action does because of the (normal) time it
takes to exchange
optical phonons of unit frequency.  The 3-dimensional version
of $S_1$ ( Eq. (\ref{eff action pot''}) ) does contain a retardation,
however, it is not just proportional to the proper time difference.
\end{description}
To maintain explicit covariance we will not
use the form (\ref{eff action pot''}). It is of course also possible
to fully perform the 4-dimensional q-integration and to obtain
\begin{equation}
S_1[x(\tau)] =
- \frac{g^2}{8 \pi^2} \int_0^{\beta} d \tau_1 \> \int_0^{\beta}
d \tau_2 \>
\frac{m}{y(\tau_1,\tau_2)} \> K_1 \left ( \> m y(\tau_1,\tau_2) \>
 \right )
\label{eff action pot'''}
\end{equation}
where $K_1(x)$ is a modified Bessel function \cite{Handbook} and
\begin{equation}
y(\tau_1,\tau_2) = \sqrt {\left [ \> x(\tau_1) \> - \> x(\tau_2) \>
\right ]^2} \> .
\end{equation}
For small relative times Eq. (\ref{eff action pot'''})
exhibits a stronger divergence ( $1 / y^2 $ ) than in the polaron
case ( $1 / y $ )  and requires the usual
renormalizations of relativistic field theory. As the Bessel function
is difficult to handle we will not use this explicit form in the
following but rather stick to the integral representation in
Eq. (\ref{eff action pot'}).

 From the derivation presented above it should be clear how
the particle representation is generalized to $N$ nucleons
(the case $N = 2$
has been considered in Ref. \cite{NTS} neglecting self-energy and
vertex corrections): to each heavy
particle there corresponds just {\em one} trajectory. This is due to
the quenched approximation which neglects production of
heavy pairs. Therefore the nucleon number is conserved and no
splitting of heavy particle trajectories can occur.

\section{Variational Approximation on the Pole of the Two-Point
Function}
\label{sec: var on 2point pole}

\noindent
In this Section we only consider the case where no external
mesons are present,
which corresponds to simply setting the meson sources $j(x)$ to zero.
The exact two-point function (or propagator) is then given by
\begin{equation}
G_2(x) = {\rm const} \int_0^{\infty} d\beta \>
\exp \left (- \frac{\beta}{2} M_0^2 \right )
\> \int_{x(0)=0}^{x(\beta)=x} {\cal D}x(\tau) \>
\exp ( - S_0\> [ x(\tau)] - S_1\> [ x(\tau)] \> ) \> .
\label{exact 2point(x)}
\end{equation}
The normalization constant can be determined by switching off the
interaction. In this case we know \cite{ItZu}
\begin{equation}
G_2 (p) \Biggr |_{S_1=0} = \int d^4 x \>
\exp ( \> i p\cdot x ) \> G_2 (x) \Biggr |_{S_1=0} \> = \>
\frac{1}{p^2 + M_0^2} \> .
\label{free 2point(p)}
\end{equation}
The correct normalization of Eq. (\ref{exact 2point(x)}) therefore
is
\begin{equation}
G_2(x) = \frac{1}{8 \pi^2} \int_0^{\infty} d\beta \>\frac{1}{\beta^2}
 \> \exp \left [- \frac{\beta}{2} M_0^2 - \>
\frac{x^2}{2 \beta} \> \right ] \> \frac { \int {\cal D}x \>
\exp( - S_0\> - S_1 \> ) } {\int {\cal D} x \> \exp ( - S_0\> ) }
\label{exact 2point(x) normalized}
\end{equation}
where the paths are subject to the boundary conditions
\begin{equation}
x(0) = 0 \> ,  \>\>\> x(\beta) = x \> .
\label{bound cond}
\end{equation}
Similarly, in momentum space we can write
\begin{equation}
G_2(p) = \frac{1}{2} \int_0^{\infty} d\beta \>
\exp \left [- \frac{\beta}{2} ( p^2 + M_0^2 ) \> \right ] \>
\frac{
\int d^4x \> \exp( i p \cdot x) \> \int {\cal D}x \>
\exp( - S_0 - S_1) } { \int d^4x \>\exp( i p \cdot x) \>
\int {\cal D}x \> \exp( - S_0 ) } \> .
\label{exact 2point(p) normalized}
\end{equation}

Due to the nonlinear dependence of the action (\ref{eff action pot'})
on the paths $x(\tau)$ it is, of course, impossible to do
the path integrals (\ref{exact 2point(x) normalized},
\ref{exact 2point(p) normalized})  exactly. However,
following Feynman \cite{Fey1}, it is possible to find a variational
approximation for the effective action starting from a solvable
trial action. This variational treatment is based on the decomposition
\begin{equation}
S = S_t \> + \> S \> - S_t \> = S_t \> + \Delta S
\label{decomp}
\end{equation}
and on Jensen's inequality
\begin{equation}
\left < e^{- \Delta S} \right >  \> \> \ge \> \> e^{-< \Delta S > }
\label{Jensen max}
\end{equation}
which holds for averages with normalized positive weighting functions.
If the weighting function is not positive (or even complex),
or $\Delta S$ is complex, the inequality in Eq. (\ref{Jensen max})
is replaced by a stationarity with respect to variations
\begin{equation}
\left < e^{- \Delta S} \right >  \> \> \buildrel {\rm stat} \over
\simeq \> \> e^{-< \Delta S > } \>.
\label{Jensen stat}
\end{equation}
Obviously, Minkowski proper time and/or Minkowski space-time only
allows the weaker form (\ref{Jensen stat}) to be used.
In addition to the choice of the trial action $S_t$
we also have the freedom how we define the averaging, i.e.
which coordinates we treat exactly and which only approximately via
the Jensen stationarity. To be more precise, one can
define
\begin{equation}
< \Delta S >_{S_t} \> \equiv \frac{ \int {\cal D}x(\tau) \>
\Delta S[x(\tau)] \> \> \exp (- S_t[x(\tau)] )
\> }{\int {\cal D}x(\tau) \> \exp (- S_t[x(\tau)] )}
\label{x averaging}
\end{equation}
or
\begin{equation}
\ll \Delta S \gg_{S_t} \> \equiv \frac{ \int d^4x \>
\exp( i p \cdot x)
\int {\cal D}x(\tau) \> \Delta S [x(\tau)] \> \>
\exp (- S_t[x(\tau)] ) }{ \int d^4x \> \exp( i p \cdot x)
\int {\cal D}x(\tau) \> \exp (- S_t[x(\tau)] )}
\label{p averaging} \>.
\end{equation}
In the first case, which we will call ``coordinate averaging'', one
has to do the Fourier transform with respect to the endpoint $x$
after the averaging to get the approximate
two-point function in momentum space
whereas in the latter (``momentum averaging'')
only the integral over the proper time $\beta$ still has to be
performed. This is reminiscent of the ``partial averaging''
procedure proposed by Doll et {\it al.} \cite{DCF} and employed in the
Monte-Carlo calculations of Ref. \cite{AlRo}. It is clear that
coordinate averaging usually is more accurate and that (with euclidean
proper time) Jensen's {\it inequality} (\ref{Jensen max})
can be used. On the other hand momentum averaging more directly gives
the two-point function in momentum space. We will see that
with suitable trial actions both averaging
procedures lead to identical results on the nucleon pole.

\subsection{Coordinate averaging}
\label{sec: coord av}

\noindent
Eq. (\ref{exact 2point(x) normalized}) may be written in the
following form
\begin{equation}
G_2(x) = \frac{1}{8 \pi^2} \int_0^{\infty} d\beta \>
\frac{1}{\beta^2} \>
\exp \left (- \frac{\beta}{2} M_0^2 - \> \frac{x^2}{2 \beta} \>
\right ) \left < \> e^{- S_1} \> \right >_{S_0} \>,
\label{exact 2point(x) as average}
\end{equation}
where the averaging is performed with respect to the weighting
function $ \exp (- S_0)$ :
\begin{eqnarray}
\left < \> e^{- S_1} \> \right >_{S_0} \> &\equiv&
\frac{\int {\cal D}x \> \exp(-S_0) \> \exp(-S_1)}
{\int {\cal D}x \> \exp(-S_0) \> } \nonumber \\
&=& < \exp ( S_t - S ) \> >_{S_t}  \> \frac{\int {\cal D}x \>
\exp(-S_t) } {\int {\cal D}x \> \exp(-S_0) \> } \> .
\label{average of exp-S1}
\end{eqnarray}
Here $S$ is the sum of $S_0$ and $S_1$.
Applying Jensen's inequality (Eq. (\ref{Jensen max})), we find
\begin{equation}
< \> e^{- S_1} \> >_{S_0} \> \ge
\> \exp ( - < \Delta S  >_{S_t}\> )  \> \frac{\int {\cal D}x
\exp(- S_t ) } {\int {\cal D}x \exp(-S_0) \> } \> .
\end{equation}
The various path integrals may be easily calculated in Fourier
space by parameterizing the paths as
\begin{equation}
x(\tau) = x \> \frac{\tau}{\beta} + \sum_{k=1}^{\infty}
\frac{2 \sqrt{\beta}} {k \pi} \> b_k \> \sin \left (
\frac{k \pi \tau}{\beta} \right ) \> .
\label{Fourier paramet of paths}
\end{equation}
This obviously fulfills the boundary conditions (\ref{bound cond}).
As only the ratio of path integrals appears in Eq.
(\ref{average of exp-S1})
the Jacobian from the transformation to Fourier space cancels and
the path integrals are now infinite-dimensional integrals over
the Fourier coefficients $ b_k\> $ for $ k = 1, \> ... \>
\infty$. If one writes the endpoint coordinate as
\begin{equation}
x = \> \sqrt{ 2 \beta} \> b_0
\label{endpoint bzero}
\end{equation}
then the free action is simply
\begin{equation}
S_0 = \> \sum_{k=0}^{\infty} \> b_k^2 \> .
\label{free Fourier action}
\end{equation}
The most general trial action
with which one can proceed analytically is one where the $b_k$'s
appear at most quadratically. We shall use
\begin{equation}
S_t = \> \sum_{k=0}^{\infty} A_k \> b_k^2 \;\;\;,
\label{Feynman Fourier action}
\end{equation}
with coefficients $A_k > 0 $ parameterized in various forms (see
below) or left free as variational parameters. A term like
$b_k \cdot b_0$ may also be introduced with only minor
complications, while off-diagonal terms  like $b_k \cdot b_{k'}$
would require the calculation of infinite-dimensional
determinants.

By this choice all path integrals are simple gaussian integrals
and can easily be performed. We obtain
\begin{equation}
\frac{\int{\cal D}{b} \> \exp(-S_t) }{\int{\cal D}{b} \>
\exp(-S_0) } \> = \> e^{ - (A_0 - 1) \> b_0^2 } \> \>
\prod_{k=1}^{\infty} \left ( \frac{1}{A_k^2} \right) \> ,
\label{Fourier path int SF over S0}
\end{equation}
\begin{equation}
< \> S_0 - S_t \> >_{S_t} = \> ( 1 - A_0) \> b_0^2 + \> 2
\sum_{k=1}^{\infty} \left ( \frac{1}{A_k} - 1 \right ),
\label{Fourier average S0-SF}
\end{equation}
and
\begin{equation}
< \> S_1 \> >_{S_t} = \> - \frac{g^2}{2} \> \int_0^{\beta} d\tau_1
d\tau_2 \> \int \frac{d^4q}{ (2 \pi)^4} \> \frac{1}{q^2 + m^2} \>
< \> \exp\left[ \> i q \cdot( x(\tau_1) - x(\tau_2) ) \> \right ]
\> >_{S_t}
\label{Fourier average S1} \> .
\end{equation}
The last average also involves a (shifted)
gaussian integral and is given by
\begin{equation}
< \> \exp\left[ \> i q \cdot( x(\tau_1) - x(\tau_2) ) \> \right ]
\> >_{S_t}
= \exp\left ( i \frac{\tau_1-\tau_2}{\beta} \> q \cdot x \> - \>
\frac{1}{2} \> \mu^2(\tau_1,\tau_2) \> q^2 \right )
\label{Fourier average exp iqx}
\end{equation}
where we have defined
\begin{equation}
\mu^2(\tau_1,\tau_2) \> = \beta \> \sum_{k=1}^{\infty}
\frac{\lambda_k^2(\tau_1,\tau_2)}{A_k}
\label{amu2}
\end{equation}
and
\begin{equation}
\lambda_k(\tau_1,\tau_2) = \frac{\sqrt{2}}{k \pi} \left [
\sin \left(\frac{k \pi \tau_1}{\beta} \right ) \> - \>
\sin \left(\frac{k \pi \tau_2}{\beta} \right ) \> \right ] \> .
\label{lambda k}
\end{equation}
We shall postpone a discussion of the meaning of the quantity
$ \mu^2 $, which plays a crucial role in what follows, until
later. Finally the $q$-integration in (\ref{Fourier average S1})
can be performed by using the representation
\begin{equation}
\frac{1}{q^2 + m^2} = \frac{1}{2} \> \int_0^{\infty} du \>
\exp \left [ - \frac{u}{2} \> (q^2 + m^2) \right ] \> .
\end{equation}
This gives
\begin{eqnarray}
< \> S_1 \> >_{S_t} = \> - \frac{g^2}{8 \pi^2} \> \int_0^{\beta}
d\sigma \int_{\sigma/2}^{\beta - \sigma/2} &dT& \int_0^{\infty}
du \> \frac{1}{ (u + \mu^2(\sigma,T)\>)^2} \nonumber \\ &\cdot&
\exp \left [ - \frac{u}{2} m^2 \> - \> \frac{x^2}{2 \beta^2}
\frac {\sigma^2}{u + \mu^2(\sigma,T)} \right ]
\label{Fourier average S1 explicit}
\end{eqnarray}
where we have used the symmetry of the integrand to restrict the
proper time integrations to $\tau_2 \le \tau_1 $ and introduced
relative and total times
\begin{equation}
\sigma = \tau_1 - \tau_2\> , \>\>\> T = \frac{1}{2} (\tau_1 +
\tau_2 ) \>.
\label{rel and tot prop time}
\end{equation}
The interaction term can be brought into simpler form
by the transformation $ u \to \mu^2/(u + \mu^2) $ which
leads to
\begin{equation}
< \> S_1 \> >_{S_t} = \> - \frac{g^2}{8 \pi^2} \> \int_0^{\beta}
d\sigma
\int_{\sigma/2}^{\beta - \sigma/2} dT \> \frac{1}{\mu^2(\sigma,T)}
\int_0^1 du \> \> e\left ( m \mu(\sigma,T),\>
\frac{x \sigma}{ \beta \mu(\sigma,T)}\> , \> u \right ) \>.
\label{Fourier average S1 explicit'}
\end{equation}
Here the function $e(s,t,u)$ is defined as
\begin{equation}
e(s,t,u) =
\exp \left ( -  \> \frac{s^2}{2} \> \frac{1-u}{u} \> - \>
\frac{t^2}{2}\> u \> \right ) \> .
\label{e(s,t,u)}
\end{equation}
In principle,
the $u$-integral can be expressed in terms of a particular plasma
dispersion function, the so-called Shkarofsky function \cite{Rob},
but there is no advantage of using this representation.

Hence, using Jensen's inequality and the trial action
(\ref{Feynman Fourier action}),
the Green function in coordinate space is bounded by
\begin{equation}
G_2(x)\> \ge \> \frac{1}{8 \pi^2} \int_0^{\infty} d\beta \>
\frac{1}{\beta^2} \>
\exp \left (- \frac{\beta}{2} M_0^2 - \> \frac{x^2}{2 \beta} \>
\right )
\> \exp \left [ - \> \beta \> \Omega(\beta) \> - \> < S_1 >_{S_t}
\right]
\label{bound for 2point(x)}
\end{equation}
where
\begin{equation}
\Omega(\beta) = \frac{2}{\beta} \> \sum_{k=1}^{\infty} \left [ \>
\ln A_k + \frac{1}{A_k} - 1 \> \right ] \> .
\label{Omega(beta)}
\end{equation}

\subsection{Renormalization}
\label{sec: renorm}

\noindent
Actually as it stands Eq. (\ref{Fourier average S1 explicit'}) does
not exist, since for small relative times (as we shall see later)
\begin{equation}
\mu^2(\sigma,T) \> \> \buildrel \sigma \to 0
\over \longrightarrow \>\> \sigma \> ,
\label{amu2 for small sigma}
\end{equation}
causing a logarithmic divergence in the
$\sigma$-integration\footnote{In $D$ dimensions the integrand
behaves like $\sigma^{D/2-1}$ which in $D = 3$ leads to the
integrable singularity $1/\sqrt{\sigma}$ of the polaron problem.}.
This is, of course, one of the
expected divergences of field theory which require renormalization.
In the present case, renormalization is particularly easy,
since only a {\em mass renormalization} for the heavy particle
is needed. In fact, the theory is
super-renormalizable in the quenched approximation -- only the
second-order
self-energy diagram of the nucleon introduces a divergence. We
regulate this with a Pauli-Villars regularization.
This amounts to subtracting a term with the meson mass replaced
by a cut-off
mass $\Lambda$ (which will eventually tend to  infinity), thus
removing the small $\sigma$-singularity. To be specific, we subtract
\begin{equation}
\frac{1}{\sigma}\>
e\left ( \> \Lambda \sqrt{\sigma}, \sqrt{\sigma} \mu_0, u \right)
\label{Pauli Villars subtraction}
\end{equation}
from $< S_1 >$, where $\mu_0$ is an arbitrary mass (renormalization
point). Since
\begin{equation}
\frac{\partial}{\partial \mu_0^2}
\frac{1}{\sigma}\>
e\left ( \> \Lambda \sqrt{\sigma}, \sqrt{\sigma} \mu_0, u \right)
\> = \> - \> \frac{u}{2} \>
e\left ( \> \Lambda \sqrt{\sigma}, \sqrt{\sigma} \mu_0, u \right)
\end{equation}
is finite at $\sigma = 0$ and vanishes for $\Lambda \to \infty$
the averaged action will be independent of $\mu_0$.
We will assume a nonzero meson mass $m$ in most of the following and
therefore the most convenient choice for us is $\mu_0 = 0$.
As shown in the Appendix one then obtains
\begin{equation}
< \> S_1 \> >_{S_t} = - \frac{g^2}{8 \pi^2} \> \beta \> \ln
\frac{\Lambda^2}{m^2} \> + \> < \> S_1 \> >^{\rm fin} \> + \> < \>
S_1 \> >^{\rm reg}\;\;\;,
\label{regularization of action}
\end{equation}
where $< \> S_1 \> >^{\rm fin} \> $ is the finite part resulting from
 the subtraction (\ref{Pauli Villars subtraction}) and is given in
Eq. (\ref{S1 finite}). The regular part reads
\begin{eqnarray}
< \> S_1 \> >^{\rm reg} = - \frac{g^2}{8 \pi^2} \> \int_0^{\beta}
d\sigma \> \int_{\sigma/2}^{\beta-\sigma/2} dT \> \int_0^1 du
&\Biggl [& \frac{1}{\mu^2(\sigma,T)} \> e\left ( m \mu(\sigma,T),\>
\frac{x \sigma}{ \beta \mu(\sigma,T)}\>, \> u \right ) \nonumber \\
&-& \frac{1}{\sigma} \> e\left ( m \sqrt{\sigma},0,u \right )
\Biggr ] \> .
\label{S1 regular}
\end{eqnarray}
 From Eq. (\ref{regularization of action}) and Eq.
(\ref{bound for 2point(x)}) it is
evident that the divergent part of the averaged action can be
absorbed into  a new mass parameter
\begin{equation}
M_1^2 = M_0^2 \> - \> \frac{g^2}{4 \pi^2} \> \ln \frac{\Lambda^2}
{m^2}
\label{finite mass}
\end{equation}
which will be found to be {\em finite}. After the bare mass has been
replaced by $M_1$ all quantities are now
well defined. Note that the renormalization (\ref{finite mass}) is
in fact the same as in lowest order perturbation theory, even though
the calculation has been done in a non-perturbative way.  Note also
that $M_1$ is in general not yet
the physical mass of the nucleon but an intermediate mass scale with
no direct physical meaning. Again, the finite shift from $M_1$ to
$M_{\rm phys}$ will be done in a non-perturbative way.

\subsection{On-mass-shell limit}
\label{sec: on mass shell}

\noindent
The physical mass is determined from the requirement that
in momentum space the two-point function  develops a pole when
approaching $ p^2 = - M_{\rm phys}^2$ :
\begin{equation}
G_2(p) \longrightarrow
\>\> \frac{Z}{p^2 + M_{\rm phys}^2} \> .
\label{pole of 2point(p)}
\end{equation}
Here $\> 0 < Z < 1 \> $ is the residue at the pole.
How is it possible that
\begin{equation}
G_2(p) = \int d^4x \> e^{ i p \cdot x} \> G_2(x) \> =
\frac{4 \pi^2}{p}
\> \int_0^{\infty} dx \> x^2 \> J_1(p x) \> G_2(x)
\end{equation}
diverges at $p = i M_{\rm phys}$ ? Obviously this can only be the
case if the large-$x$ behaviour of $G_2(x)$ (which is only
a function of $x^2$ ) is not able to overcome
the exponential growth of the Bessel function \cite{Handbook}
\begin{equation}
J_1\left (i M_{\rm phys} \> x \right) = i \> I_1
\left ( M_{\rm phys} x \right) \> \>
\buildrel x\to \infty \over \longrightarrow \> \> i \>
\frac{e^{M_{\rm phys} \> x}}{\sqrt{2 \pi M_{\rm phys} \> x}} \>.
\end{equation}
Therefore   the physical mass is given by
\begin{equation}
M_{\rm phys} = - \lim_{x \to \infty} \> \frac{1}{x} \> \ln \,
( \>G_2(x)\> )\;\;\;.
\label{Mphys from large x}
\end{equation}
This is similar to the way the ground-state energy is obtained from
the partition function in non-relativistic physics or
the mass of hadrons in lattice calculations.

However, the explicit expression for $G_2(x)$
(\ref{bound for 2point(x)})
contains a term $\exp( - x^2/2\beta) $ which would decay like a
gaussian unless the proper time $\beta$ is proportional to $x$
and also tends to infinity. These heuristic arguments suggest
that we have to study the limit $ x, \beta \to \infty$ but keep
\begin{equation}
\lambda = \frac{1}{M_{\rm phys}} \> \frac{x}{\beta}
\label{def lambda}
\end{equation}
fixed. In Eq. (\ref{def lambda}) the extra factor
$M_{\rm phys}^{-1}$ has been introduced to obtain a
dimensionless quantity~\footnote{Recall from
Eq. (\ref{euclid proper time}) that our proper time
has dimension (mass)$^{-2}$ . }.
 From Eq. (\ref{bound for 2point(x)}) we then obtain
\begin{equation}
G_2(x)  \ge \> \frac{1}{8 \pi^2} \> \frac{M_{\rm phys}}{x} \>
\int_0^{\infty} d\lambda \> \> e^{\> - x \> F(x,\lambda) }
\label{G2 lambda}
\end{equation}
where
\begin{equation}
F(x,\lambda) = \frac{M_1^2}{2 \lambda M_{\rm phys}} +
\frac{\lambda}{2} M_{\rm phys} + \frac{1}{\lambda M_{\rm phys}}\>
\Omega \> + \> \frac{1}{x} \left ( \> < \> S_1 \> >^{\rm fin}
\> + \> < \> S_1 \> >^{\rm reg} \right )\>.
\end{equation}
In the limit $x \to \infty$ Laplace's method  \cite{Sir,Hen} tells
us that Eq. (\ref{G2 lambda}) behaves like
\begin{equation}
G_2(x) \> \>\buildrel x \to \infty \over  \ge
 \> \> \frac{{\rm const}}{x^{3/2}} \> e^{ \> - x \> F(\lambda_0) }
\end{equation}
where $F(\lambda_0)$ is the minimum of $F(x \to \infty,\lambda)$ .
Inserting this result into Eq. (\ref{Mphys from large x}) we
obtain
\begin{equation}
M_{\rm phys} \le F(\lambda_0) \>.
\label{Mphys from F(lambda)}
\end{equation}
We have to study the large $x$- and the large $\beta$-limit of the
averaged action. First, we note from Eq. (\ref{S1 finite}) that
for $\mu_0 = 0$
\begin{equation}
\lim_{\beta \to \infty} \> \frac{1}{\beta} < \> S_1 \> >^{\rm fin}
\> \> = \> 0.
\end{equation}
Then we assume that
\begin{equation}
\lim_{\beta \to \infty} \>\mu^2(\sigma,T) = \mu^2(\sigma)
\label{amu2 large beta}
\end{equation}
which holds in all parametrizations which we will study. Therefore
\begin{eqnarray}
V \> \equiv \>
\lim_{\beta \to \infty} \> \frac{1}{\beta} < \> S_1 \> >^{\rm reg}
= \> &-& \> \frac{g^2}{8 \pi^2} \> \int_0^{\infty} d\sigma\>
\int_0^1 du \> \Biggl [ \> \frac{1}{\mu^2(\sigma)} \nonumber \\
&\cdot& e \> \left ( m \mu(\sigma),
\frac{\lambda M_{\rm phys} \sigma}{ \mu(\sigma)}, u \right)
\> - \> \frac{1}{\sigma} \> e \> ( m \sqrt{\sigma},0,u)
\> \Biggr ]
\label{pot}
\end{eqnarray}
has a well-defined limit. We will also assume (and later verify)
that $\Omega(\beta)$ defined in Eq. (\ref{Omega(beta)}) has a
large-$\beta$ limit
\begin{equation}
\Omega \> = \> \lim_{\beta \to \infty} \> \Omega(\beta) \> .
\label{Omega}
\end{equation}
Suppressing the subscript zero for $\lambda$ we finally arrive at
the following inequality for the physical mass
\begin{equation}
M_{\rm phys}^2 \le \> \frac{M_1^2}{2 \lambda} \> + \>
\frac{\lambda}{2} \> M_{\rm phys}^2 \> + \> \frac{1}{\lambda}
\> \left (\Omega \> + \> V \right )\>.
\label{var inequality for Mphys}
\end{equation}
Eq. (\ref{var inequality for Mphys}) is the main result of this
Section. Since $M_{\rm phys}$ is fixed we can turn it around and use
\begin{equation}
M_1^2 \> \ge \> (2 \lambda - \lambda^2 ) \> M_{\rm phys}^2 -
2 \> \left (\Omega \> + \> V \right )
\label{var inequality for M1}
\end{equation}
to maximize the r.h.s with respect to $\lambda$ and all the
parameters in the trial action. In the following we will call
$\Omega$ the ``kinetic'' term
because it has no explicit coupling constant dependence and $ V $
the ``potential'' term because it has. In addition,
Eq. (\ref{var inequality for Mphys}) looks like a variational
equation for the energy in nonrelativistic quantum mechanics.

{\em Without} variation the equality sign in
Eq. (\ref{var inequality for M1}) gives the perturbative result
from the one-loop graph shown in Fig. 3 (a). This can be seen
as follows: while we expect $\lambda = 1 + {\cal O}(g^2) $
the combination
$2 \lambda - \lambda^2$ is $1 + {\cal O}(g^4) $. Similarly, from
$A_k = 1 + {\cal O}(g^2) $, we deduce $\Omega = {\cal O}(g^4) $ (see
Eq. (\ref{Omega(beta)}) ) and $\mu^2(\sigma) = \sigma +
{\cal O}(g^2)$. Therefore to lowest order in $g^2$ we obtain
\begin{equation}
M_1^2 \> = \> M_{\rm phys}^2 \> - \> 2 \> V
\Bigr |^{\lambda=1}_{\mu^2(\sigma)=\sigma} \> + {\cal O}(g^4)\;\;\;,
\end{equation}
or
\begin{equation}
M_{\rm phys}^2 = M_1^2 \> + \> \frac{g^2}{4 \pi^2} \> \int_0^1 du \>
\ln \left [ 1 + \frac{M_{\rm phys}^2}{m^2} \frac{u^2}{1-u} \> \right ]
\label{perturb result from var}
\end{equation}
after performing the $\sigma$-integral. The same result is obtained
from the direct calculation of the self-energy diagram in Fig. 3 (a)
\begin{equation}
\Sigma(p^2) \> = \> - \frac{g^2}{4 \pi^2} \> \ln \frac{\Lambda^2}{m^2}
\> + \> \frac{g^2}{4 \pi^2} \> \int_0^1 du \>
\ln \left [ 1 + \frac{p^2}{m^2} u + \frac{M_0^2}{m^2} \frac{u}{1-u} \>
\right ] \> .
\label{perturb self energy}
\end{equation}
The pole position is determined by $M_{\rm phys}^2 = M_0^2 +
\Sigma( - M_{\rm phys}^2) $ from which we obtain Eq.
(\ref{perturb result from var}) in lowest order after renormalizing
the mass (see Eq. (\ref{finite mass})).

\begin{figure}
\unitlength=1mm
\begin{picture}(33,35)
\put(0,5){\makebox(33,35){\psfig{figure=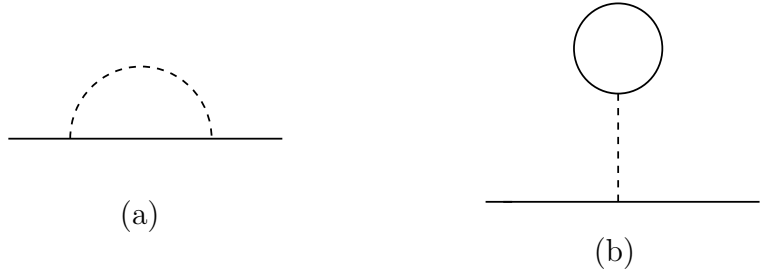,height=200mm,width=500mm}}}
\put(56,5){(a)}
\put(119,0){(b)}
\end{picture}
\caption{Second-order graphs for the two-point function:
(a) self-energy graph, (b): tadpole graph. In the quenched approximation
the tadpole graph is neglected}
\end{figure}

\subsection{Momentum averaging}
\label{sec: mom av}

\noindent
In coordinate averaging the determination of the physical mass
was a rather involved
procedure. This is avoided in ``momentum averaging'', where
we also average over the endpoint coordinate $x$ with the additional
weight $\exp (i p \cdot x)$. This extra weight can be formally
absorbed in a modified (complex) free action
\begin{equation}
\tilde S_0 = S_0 - i p \cdot x \>.
\label{tilde free action}
\end{equation}
In other words, we write Eq. (\ref{exact 2point(p) normalized}) as
\begin{equation}
G_2(p) \> = \> \frac{1}{2} \> \int_0^{\infty} d\beta \>
\exp \left [ - \frac{\beta}{2} ( p^2 + M_0^2)  \right ] \> \cdot \>
\ll \> e^{-S_1} \> \gg_{\tilde S_0}
\end{equation}
where
\begin{eqnarray}
\ll \> e^{- S_1} \> \gg_{\tilde S_0} \> &\equiv&
\frac{\int {\cal D}\tilde x
\> \exp(-\tilde S_0) \> \exp(-S_1)}{\int {\cal D}\tilde  x \>
\exp(-\tilde S_0) \> } \nonumber \\
&=& \ll \exp ( \tilde S_t - \tilde S ) \> \gg_{\tilde S_t}  \>
\frac{\int {\cal D}\tilde x \> \exp(-\tilde S_t) }
{\int {\cal D}\tilde x \> \exp(-\tilde S_0) \> } \> .
\label{tilde average of exp-S1}
\end{eqnarray}
Here we have defined
\begin{equation}
\int {\cal D}\tilde  x \> ... \> = \int d^4x \> \int {\cal D} x \> ...\;\;\;.
\end{equation}
Because the weight function is now complex, we can only apply
Jensen's stationarity relation (\ref{Jensen stat})
\begin{equation}
\ll \> e^{- S_1} \> \gg_{\tilde S_0} \> \> \simeq  \>
\exp \left ( - \ll S - \tilde S_t \gg_{\tilde S_t} \> \right )
\> \frac{\int {\cal D}\tilde x \> \exp(-\tilde S_t) }
{\int {\cal D}\tilde x \> \exp(-\tilde S_0) \> } \> .
\end{equation}
As trial action we take
\begin{equation}
\tilde S_t = \sum_{k=0}^{\infty} A_k \> b_k^2 \> -
\> i \> \tilde \lambda \> p \cdot x
\label{tilde trial action}
\end{equation}
where $\tilde \lambda$ is an additional variational parameter which
rescales the momentum.

As the evaluation of the various path integrals closely follows the
one in Section \ref{sec: coord av}
we can be brief and just state the results
\begin{equation}
\frac{\int {\cal D}\tilde x \> \exp(-\tilde S_t) }
{\int {\cal D}\tilde x \> \exp(-\tilde S_0) \> }
= \> \exp \left ( - \frac{\beta}{2} p^2 ( \frac{\tilde \lambda^2}
{A_0} - 1) \> \right ) \> \> \prod_{k=0}^{\infty} \left (
\frac{1}{A_k^2} \right),
\label{tilde Fourier path int SF over S0}
\end{equation}
\begin{equation}
\ll \> \tilde S_0 - \tilde S_t \> \gg_{\tilde S_t} =
\> 2 \sum_{k=0}^{\infty}
\left ( \frac{1}{A_k} - 1 \right ) \> - \> \frac{\beta}{2} p^2 \>
\frac{ \tilde \lambda}{A_0^2} (\tilde \lambda + \tilde \lambda
A_0 - 2 A_0 ) \> ,
\label{tilde Fourier average S0-SF}
\end{equation}
and this time the interaction term is
\begin{equation}
\ll \> S_1 \> \gg_{\tilde S_t} = \> - \frac{g^2}{8 \pi^2}
\> \int_0^{\beta} d\sigma
\int_{\sigma/2}^{\beta - \sigma/2} dT \> \frac{1}{\tilde
\mu^2(\sigma,T)} \int_0^1 du \> \> e\left ( m \tilde \mu(\sigma,T),
\> \frac{- i \tilde \lambda p \sigma}{ A_0 \tilde \mu(\sigma,T)}
\> , \> u \right ) \>.
\label{tilde Fourier average S1 explicit'}
\end{equation}
Here
\begin{equation}
\tilde \mu^2(\sigma,T) \> = \>  \frac{\sigma^2}{A_0 \beta} \>
+ \> \mu^2(\sigma,T) \> .
\label{tilde amu2}
\end{equation}
Renormalization of the averaged action is along the same lines as
in the Appendix. Combining all terms we obtain the propagator in
momentum space
\begin{eqnarray}
G_2(p)\> &\simeq& \> \frac{1}{2} \int_0^{\infty} d\beta \>
\exp \left (- \frac{\beta}{2} ( p^2 + M_1^2 ) + \frac{\beta}{2}
p^2 (1 - \frac{\tilde \lambda}{A_0})^2
\> \right ) \nonumber \\
&\cdot& \exp \left ( - \> \beta \> \tilde \Omega(\beta) \> -
\> \ll S_1 \gg^{\rm reg}
- \> < S_1 \> >^{\rm fin} \right)
\label{var approx for 2point(p)}
\end{eqnarray}
where
\begin{equation}
\tilde \Omega (\beta) = \frac{2}{\beta} \> \sum_{k=0}^{\infty}
\left [ \> \ln A_k + \frac{1}{A_k} - 1 \> \right ]
\label{tilde Omega(beta)}
\end{equation}
and
\begin{eqnarray}
\ll \> S_1 \> \gg^{\rm reg} = - \frac{g^2}{8 \pi^2} \>
\int_0^{\beta}
d\sigma \> \int_{\sigma/2}^{\beta-\sigma/2} dT \> \int_0^1 du
&\Biggl [& \frac{1}{\tilde \mu^2(\sigma,T)} \>
e\left ( m \tilde \mu(\sigma,T),\>
\frac{- i \tilde \lambda p \sigma}{ A_0 \tilde \mu(\sigma,T)}\>,
\> u \right ) \nonumber \\
&-& \frac{1}{\sigma} \> e\left ( m \sqrt{\sigma},0,u \right )
\Biggr ] \> .
\label{tilde S1 regular}
\end{eqnarray}
Because the small $\sigma$-behaviour of $\tilde \mu^2(\sigma,T)$
is the same as that of $\mu^2(\sigma,T)$ (see Eq.
(\ref{tilde amu2}) ) we have subtracted the
same term (\ref{Pauli Villars subtraction}) as before. This explains
 why the finite part $< S_1 >^{\rm fin}$ of the averaged action is
unchanged.

The on-shell limit of Eq. (\ref{var approx for 2point(p)})
is now particularly easy : a pole develops if in
\begin{equation}
G_2(p) \simeq \frac{1}{2} \> \int_0^{\infty} d\beta \>
\exp \left [ -\frac{\beta}{2} F(\beta,p^2) \right ]
\label{2point func as beta int over F}
\end{equation}
the function $F\left (\beta \to \infty, p^2 = - M_{\rm phys}^2
\right) $ vanishes. This leads to
\begin{equation}
M_{\rm phys}^2 = \>  M_1^2  + M_{\rm phys}^2 \> \left ( 1 -
\frac{\tilde \lambda}{A_0} \right )^2 + 2 \lim_{\beta \to \infty}
\> \left [ \tilde \Omega(\beta) + \frac{1}{\beta}
\ll S_1 \gg^{\rm reg}
\Biggr |_{p=i M_{\rm phys}} \> \right ] \> .
\label{stat eq for Mphys}
\end{equation}
For any sensible parametrization $A_0$ is finite in the large
$\beta$-limit. Therefore
the tilde can be dropped from $\tilde \mu^2(\sigma)$ and
$\tilde \Omega$ for large $\beta$ (see Eqs.
(\ref{tilde amu2}, \ref{tilde Omega(beta)}) ) and Eq.
(\ref{stat eq for Mphys})
is completely equivalent to Eq. (\ref{var inequality for Mphys})
if we identify
\begin{equation}
\tilde \lambda = A_0 \> \lambda \> .
\end{equation}
Due to the use of a complex trial action momentum averaging only
tells us that the r.h.s. of Eq. (\ref{stat eq for Mphys}) is an
extremum (and not necessarily a minimum) under variations.
Since the intermediate mass scale $M_1$ does not show up in any
observables this has no direct physical consequences. Of course,
a minimum principle has the extra advantage that the minimal
value gives a clear measure of the quality of the variational ansatz.

\section{Variational Ans\"atze}
\label{sec: var ansaetze}

\noindent
Having developed the general formalism for the variational
calculation in the last two sections we now need to turn our
attention to the specific form of the trial action
(\ref{Feynman Fourier action}). We shall first consider
two specific parametrizations of the Fourier coefficients $A_k$ of
this action, followed by the best possible parameterization
(within the gaussian ansatz) where the actual functional form of
the $A_k$'s is determined by the variational principle. Before we
do this, however, it is useful to discuss some general features of
the trial action. We begin by writing down the general quadratic
two-time action in coordinate space
\begin{equation}
S_t[x] \> = \> \int_0^{\beta} d\tau \> \frac{1}{2} \dot x^2 \> + \>
\int_0^{\beta} d\tau_1 \> \int_0^{\tau_1} d\tau_2 \>
f ( \tau_1 - \tau_2 ) \> \left [ \> x(\tau_1) \> - \> x(\tau_2) \>
\right ]^2
\label{general x Feynman action}
\end{equation}
where $f ( \tau_1 - \tau_2 ) $ is an undetermined
{\em retardation function}. Inserting the Fourier parametrization
(\ref{Fourier paramet of paths}) of the paths we obtain the
following expressions for the Fourier coefficients $A_k$
\begin{eqnarray}
A_0 (\beta) &=& \> 1 \> + \> 2 \int_0^{\beta} d\sigma \> f(\sigma)
\> \sigma^2 \> \left ( \> 1 \> - \> \frac{\sigma}{\beta} \right )
\label{A0 full}\\
A_k(\beta) &=& \> 1 \> + \> \frac{8 \beta^2}{k^2 \pi^2} \>
\int_0^{\beta}
d\sigma \> f(\sigma) \>  \left ( \> 1 \> - \> \frac{\sigma}{\beta} -
\frac{1}{k \pi}
\sin \frac{k \pi \sigma}{\beta}
\right ) \> \sin^2 \frac{k \pi \sigma}{2 \beta} \> \> \> \>,
k = 1, 2  \> ...
\label{Ak full}
\end{eqnarray}
Here we have neglected cross terms  of the form
$\> \> b_k \cdot b_{k'} \> \> \> \> , \> k, k' = 0,1 \> ...$  which
are suppressed for large $\beta$ \cite{AlRo}. It is therefore
consistent to also take the large $\beta$-limit of Eqs.
(\ref{A0 full}, \ref{Ak full}). This gives
\begin{equation}
A_k(\beta) =\>  1 \> + \> \frac{8 \beta^2}{k^2 \pi^2} \>
\int_0^{\infty}
d\sigma \> f(\sigma) \>  \sin^2 \frac{k \pi \sigma}{ 2 \beta}\> ,
\> \>\>\> k = 0,1 \> ...
\label{Ak large beta}
\end{equation}
In the following we will use only this form. Note that in this
expression the dependence on $\beta$ and the number $k$ of the
Fourier mode only comes in via the combination
\begin{equation}
E \> = \> \frac{k \pi}{\beta}
\label{def E} \>.
\end{equation}
Writing $ A_k \equiv A (k \pi/\beta) $, in particular
$ A_0 = A(0) $ , we therefore have
\begin{equation}
A( E ) =\>  1 \> + \> \frac{8}{E^2} \> \int_0^{\infty}
d\sigma \> f(\sigma) \>  \sin^2 \frac{E \sigma}{ 2 } \> .
\label{A(E)}
\end{equation}
Clearly $ A(E) $ is even :
\begin{equation}
A(- E ) \> = \> A( E )
\label{A(E) is even}
\end{equation}
and tends to unity for large $E$
\begin{equation}
A (E) \buildrel E \to \infty \over \longrightarrow \> \> 1 \>.
\label{A(E=infty)}
\end{equation}
The way how this limit is approached depends on the small-
$\sigma$ behaviour
of the retardation function $f(\sigma)$. We should emphasize that
the trial action which we use is given by
\begin{equation}
S_t = \sum_{k=0}^{\infty} \> A\left(\frac{k \pi}{\beta} \right ) \>
b_k^2
\end{equation}
in Fourier space and not by Eq. (\ref{general x Feynman action})
in $x$-space. However, since one usually
has more intuition in coordinate space it is useful to deduce
general properties and special parametrizations for the ``profile
function'' $A(E)$ from the $x$-space formulation.

We are now in a position to express the quantities $\mu^2(\sigma)$
and $\Omega$ in terms of $A(E)$. The tool to perform the sums over
Fourier modes in Eqs. (\ref{amu2}, \ref{Omega} )
is Poisson's summation formula \cite{Light,Act}
\begin{equation}
\sum_{k=-\infty}^{+\infty} \> F( k )\> = \>
\sum_{n=-\infty}^{+\infty} \> \int_{-\infty}^{+\infty} dx \> F ( x )
\> e^{2 i \pi n x}
\label{Poisson sum}
\end{equation}
which, for an even function $F(k \pi/\beta)$, leads to
\begin{equation}
\sum_{k=1}^{\infty} \> F \left ( \frac{k \pi}{\beta} \right )\> = \>
\frac{\beta}{\pi} \int_0^{\infty} dE \> F( E ) - \frac{1}{2} F( 0 )
+ \frac{2 \beta}{\pi}
\sum_{n=1}^{\infty} \> \int_0^{\infty} dE \> F ( E )\>
\cos ( 2 n \beta E ) \>.
\label{Poisson even}
\end{equation}
This is an exact form which is much more useful for our purposes
than, for example, the Euler-MacLaurin summation formula
\cite{Handbook}. The usefulness of Eq. (\ref{Poisson even}) comes
from the fact that for ordinary functions the asymptotic behaviour
of the Fourier cosine transformation \cite{Light} is given by
\begin{equation}
\int_0^{\infty} dx \> F(x) \cos \> (2 x y ) \> \sim \> -
\frac{F'(0)}
{(2 y)^2} \> + \> \frac{F'''(0)}{(2 y)^4} \> - \> ...
\label{Fourier cos asymp}
\end{equation}
Since $A(E)$ is even all odd derivatives at $E = 0$ will vanish,
unless $F(x)$ is singular at $x = 0$.
Therefore the asymptotic fall-off of the last term in
Eq. (\ref{Poisson even}) with increasing $\beta$
will not be powerlike but in most cases at least exponential. For
brevity such terms will be denoted by
\begin{equation}
{\rm Ex}_i \> (\beta) \> \equiv \>
 \frac{2 \beta}{\pi}
\sum_{n=1}^{\infty} \> \int_0^{\infty} dE \> F ( E )\>
\cos ( 2 n \beta E)  \>,
\label{Poisson exponential remainder}
\end{equation}
where $i$ is an index with which we label the various functions
$ F $ which occur. Let us first apply Poisson's summation formula
(\ref{Poisson even}) to the sum in Eq. (\ref{amu2}).
Recalling the definitions (\ref{lambda k}) and
(\ref{rel and tot prop time}) we obtain
\begin{eqnarray}
\mu^2(\sigma,T) &=& \> 8 \> \beta \> \sum_{k=1}^{\infty} \>
\frac{1}{A_k} \frac{1}{k^2 \pi^2} \> \sin^2
\frac{k \pi \sigma}{2 \beta}
\cos^2 \frac{k \pi T}{ \beta} \nonumber \\
&=& \frac{8}{\pi} \int_0^{\infty} dE \> \frac{1}{A(E)}
\frac{1}{E^2} \>
\sin^2 \frac{E \sigma}{2} \> \cos^2 E T  \> - \>
\frac{\sigma^2}{\beta A(0)} \> + \> {\rm Ex}_1\> (\beta)  \> .
\label{amu2(sigma,T) Poisson}
\end{eqnarray}
The trigonometric identity $ \cos^2 E T = (1 + \cos 2 E T )/2 $
allows us to simplify Eq. (\ref{amu2(sigma,T) Poisson}) further:
again the cosine term only contributes to exponentially small
terms \footnote{Strictly speaking these terms are exponentially
small in $T$, not $\beta$. In order to obtain sensible asymptotic
behaviour for the theory, however, it is necessary for the trial
action (\ref{general x Feynman action}) to receive its main
contribution for $\tau_{1,2}$ not too close to the endpoints of
the path. Hence $ T = (\tau_1 + \tau_2)/2 $ must grow
like $\beta$ .} so that
\begin{equation}
\mu^2(\sigma,T) = \> \frac{4}{\pi} \int_0^{\infty} dE \>
\frac{1}{A(E)} \>
\frac{\sin^2 (E \sigma/ 2)}{E^2} \> - \> \frac{\sigma^2}{\beta A(0)}
+ {\rm Ex}_2 \>(\beta)  \> .
\label{amu2(sigma,T) Poisson'}
\end{equation}
In this form the limit $ \> \beta \to \infty \> $ is trivial and
given by the simple formula
\begin{equation}
\mu^2(\sigma) \equiv \> \lim_{\beta \to \infty} \mu^2(\sigma,T) \> =
\>  \frac{4}{\pi} \int_0^{\infty} dE \> \frac{1}{A(E)} \>
\frac{\sin^2 (E \sigma/ 2)}{E^2} \>.
\label{amu2(sigma)}
\end{equation}
We further note that because of Eq. (\ref{A(E=infty)}) the small
$\sigma$-limit of $\mu^2$ is
\begin{equation}
\lim_{\sigma \to 0} \> \mu^2(\sigma) = \>  \frac{4}{\pi}
\int_0^{\infty} dE \>
\frac{\sin^2 (E \sigma/ 2)}{E^2} \> = \> \sigma\;\;\;,
\label{amu2(sigma small)}
\end{equation}
which is what we have used for discussion of the divergences in
the averaged action (see Eq. (\ref{amu2 for small sigma})).
The large-$\sigma$ limit is given by
\begin{equation}
\lim_{\sigma \to \infty} \> \mu^2(\sigma) = \>  \frac{4}{\pi}
\frac{1}{A(0)} \> \int_0^{\infty} dE \>
\frac{\sin^2 (E \sigma/ 2)}{E^2} \> = \> \frac{\sigma}{A(0)}\;\;\;.
\label{amu2(sigma large)}
\end{equation}
Because both the small and large $\sigma$ limit of $ \mu^2(\sigma)$
are proportional to $ \sigma $ we shall call it a ``pseudotime''.

We now turn to the sum over Fourier modes in Eq. (\ref{Omega(beta)}) .
By applying Eq. (\ref{Poisson even}) one easily obtains
\begin{equation}
\Omega(\beta) \> = \> \frac{2}{\pi} \> \int_0^{\infty} dE \>
\left [ \> \ln A(E) \> + \> \frac{1}{A(E)} \> - \> 1 \>
\right ] \> - \frac{1}{\beta}
\> \left [ \> \ln A(0) \> + \> \frac{1}{A(0)} \> - \> 1 \>
\right ] \> + {\rm Ex}_3 \> (\beta)
\label{Omega(beta) Poisson}
\end{equation}
so that
\begin{equation}
\Omega \> = \> \lim_{\beta \to \infty} \Omega(\beta) \>
= \> \frac{2}{\pi} \> \int_0^{\infty} dE \>
\left [ \> \ln A(E) \> + \> \frac{1}{A(E)} \> - \> 1 \> \right ] \>.
\label{Omega by A(E)}
\end{equation}
For convergence of the integral $A(E)$ has to approach unity faster
than $ 1/\sqrt{E}$ for large $E$.

\subsection{Feynman parametrization}
\label{sec: Feyn}

\noindent
In his famous polaron paper, Feynman \cite{Fey1}
chose the retardation function
\begin{equation}
f(\sigma) \equiv f_F(\sigma) \> = \> C \> e^{-w \sigma} \> ,
\label{Feynman retard func}
\end{equation}
with $C$ and $w$ as variational parameters.
This was motivated by the exact polaron effective action
(\ref{polaron action}), which has an exponential retardation function
due to the time it takes for phonons to be emitted and reabsorbed
by the electron. Furthermore,
it may be argued \cite{FeHi} that the exponential suppression at large
relative times suppresses, at least partially, the increase of the
quadratic trial action (\ref{general x Feynman action}) for large
$x(\tau_1) - x(\tau_2)$. (The exact action  obviously goes to zero
in this limit.) For this reason we will still adopt Eq.
(\ref{Feynman retard func}) for the variational approximation to
the meson-nucleon action (Eq. (\ref{eff action pot'})) in a first
try in this subsection, even though
now, of course, there is no explicit
retardation function in  {\em proper} time in this action. We
will see that this  allows many calculations to be done analytically.
In the next subsections we will consider more general trial actions.

Again following Feynman, we replace the strength $ C > 0$ by a
parameter $v$ via
\begin{equation}
v^2 \> = \>w^2 \> + \> \frac{4 C}{w} \>.
\label{Feynman v}
\end{equation}
It is obvious that $v$ has to be larger than $w$.
 From Eq. (\ref{A(E)}) we obtain
\begin{equation}
A_F ( E ) \> = \> \frac{v^2 \> + \> E^2}{w^2 \> + \> E^2}\;\;\;.
\label{Feynman A(E)}
\end{equation}
Note that as a function of the {\em complex} variable $\> E\> $
Feynman's profile function vanishes at $E = \pm i v$ which in
Minkowski space
determines the location of the caustics (or
focal points)\cite{Schul}. In addition $A_F(E)$ has poles at
$E = \pm i w$ . From Eq. (\ref{amu2(sigma)}), we obtain the pseudotime
\begin{equation}
\mu^2_F(\sigma) = \frac{w^2}{v^2} \> \sigma \> + \>
\frac{v^2 - w^2}{v^3} \>
\left ( \> 1 \> - \> e^{-v \sigma} \> \right ) \> .
\label{Feynman amu2}
\end{equation}
The limits (\ref{amu2(sigma small)}) and (\ref{amu2(sigma large)})
can be read
off directly from this explicit form. Finally one obtains
\begin{equation}
\Omega_F \> = \> \frac{ (v - w)^2}{v}
\label{Feynman Omega}
\end{equation}
which is the $D = 4$ generalization of the polaron result
\footnote{In the
polaron case the kinetic term in the variational expression for the
energy is $3 (v-w)^2/4v \> $ \cite{Fey1} .}.

\subsection{An improved retardation function}
\label{sec: improved}

\noindent
The Feynman parametrization outlined above has the advantage that
it is extremely simple and that many manipulations may be done
analytically. It has the disadvantage that for small $ \sigma $
it exhibits a different behaviour to the true action, which is
singular at this point. We shall now
indicate heuristically how one may arrive at a trial action which
does exhibit this singularity behaviour. To start off with, we
shall add a constant term to the previous action
(\ref{general x Feynman action}) :
\begin{equation}
\int_0^{\beta} d\tau_1 \> \int_0^{\tau_1} d\tau_2 \> \left [
g(\tau_1 - \tau_2)
\> + \> f(\tau_1 - \tau_2) \> \left ( x(\tau_1) - x(\tau_2)
\right )^2 \right ]\;\;\;.
\label{quadrat approx with constant}
\end{equation}
This should mimic the exact action (\ref{eff action pot'})
as much as possible. Here we have written the constant term (which
cancels in the averaging procedure) as a double-time integral over
a function
$g(\tau_1 - \tau_2)$. We can determine the functions $f$ and $g$
approximately by requiring that on the level of the proper time
integrands the momentum averaging of
(\ref{quadrat approx with constant}) should be equal to the momentum
averaging of the exact action. To avoid nonlinear equations we
perform the averaging with the free action. Using Eqs.
(\ref{tilde Fourier average S1 explicit'}) and (\ref{tilde amu2})
in the large-$\beta$ limit and setting $ \tilde \lambda = A_0 = 1 ,
\> \> \>
\mu^2 (\tau_1,\tau_2) = \tau_1 - \tau_2 = \sigma $ we obtain
\begin{equation}
g(\sigma)+ f(\sigma)  \ll \left ( x(\tau_1) -
x(\tau_2) \right )^2 \gg_{\tilde S_0} \> \> \simeq
\> - \frac{g^2}{8 \pi^2} \>  \frac{1}{\sigma}
\int_0^1 du \> \> e\left ( m \sqrt{\sigma},
- i p \> \sqrt{ \sigma}, \> u\right ) \>.
\end{equation}
If we approximate the $u$-integral by taking the integrand at some
$u =  \bar u $ we obtain
\begin{equation}
g(\sigma)+ f(\sigma) \> \ll \left ( x(\tau_1) -
x(\tau_2) \right )^2 \gg_{\tilde S_0} \> \>
\simeq - \frac{g^2}{8 \pi^2}
 \frac{1}{\sigma} \exp \left [ - \frac{1}{2} \>
( m^2 \frac{1- \bar u}{\bar u} - p^2 \bar u )\> \sigma \right ] \> .
\end{equation}
Furthermore, as a special case of the general averaging
(\ref{Fourier average exp iqx}) we have
\begin{equation}
\ll \left ( x(\tau_1) -
x(\tau_2) \right )^2 \gg_{\tilde S_0}  = 4 \sigma  - \sigma^2 p^2
\end{equation}
which is well known in Brownian motion : at small times the mean
square distance in a diffusion process grows linearly with the
time. Expanding around $ p^2 = - M^2_{\rm phys} $ and comparing
coefficients we finally obtain for the retardation function
$f(\sigma)$
\begin{eqnarray}
f_I(\sigma) &\simeq&
\frac{g^2}{32 \pi^2}
 \frac{1}{\sigma^2} \exp \left [ - \frac{1}{2} \> \left (
m^2 \frac{1 - \bar u}{\bar u}
+ M_{\rm phys}^2\bar u \right ) \> \sigma \right ] \nonumber \\
&=& \frac{C'}{\sigma^2} \> e^{ - w \sigma} \> .
\label{improved retard func}
\end{eqnarray}
The most remarkable feature of the `improved' retardation function
(\ref{improved retard func}) is that it is singular at small relative
times and thereby simulates the singular behaviour of the exact
effective action. Although Eq. (\ref{improved retard func}) gives
explicit values for the constants $C'$ and $w$ these should not be
taken too seriously as they are derived from averaging with the
free action. We will only use
the form of the retardation function as suggested
by Eq. (\ref{improved retard func}) and again treat $C'$ and $w$
as variational parameters. The resulting profile function is
\begin{equation}
A_I(E) = \> 1 + \frac{4 C'}{E} \> \left [ \arctan \frac{E}{w} -\>
\frac{w}{2 E} \ln \left ( 1 + \frac{E^2}{w^2} \right ) \> \right ]
\> .
\label{improved A(E)}
\end{equation}
At large $E$ this falls off only like
\begin{equation}
A_I(E) \buildrel E \to \infty \over \longrightarrow \> 1 \> + \>
\frac{2 \pi C'}{E} \> + \> ... ,
\label{improved A(E) for large E}
\end{equation}
which reflects the small $\sigma$-behaviour of the retardation
function. Furthermore, $A_I(E)$ now has a {\em branch point} at
$ E = \pm i w $ which will
become important when we study processes like meson production and
scattering in subsequent applications.
Again, we can eliminate the strength parameter $C'$ in terms of a
parameter $v$ by writing $A_I(0) = v^2/w^2$. This determines
\begin{equation}
C' = \frac{1}{2 w} \> ( v^2 - w^2)
\label{C' expressed by v,w} \> .
\end{equation}
We have been unable to find analytical expressions for
$\mu^2(\sigma)$ and
$\Omega$ with the profile function (\ref{improved A(E)}). They will
be calculated numerically in all the applications which follow.

\subsection{Variational equations}
\label{sec: var eq}

\noindent
The optimal choice for the retardation function is obtained if one
doesn't restrict its functional form in the way we have done in the
two cases above, but rather determines this form through
the variational principle itself. In the polaron case this approach
was first proposed by Adamowski {\it et al.} \cite{AGL} and
Saitoh \cite{Sai}. It corresponds to varying Eq.
(\ref{var inequality for M1}) with respect to
$\lambda$ and the profile function $A(E)$.
We first recall from Eq. (\ref{amu2(sigma)}) that the pseudotime
$\mu^2(\sigma)$ can be expressed through the profile function by
\begin{equation}
\mu^2(\sigma) \> =
\>  \frac{4}{\pi} \int_0^{\infty} dE \> \frac{1}{A(E)} \>
\frac{\sin^2 (E \sigma/ 2)}{E^2} \>.
\label{var eq for amu2}
\end{equation}
We may then
vary Eq. (\ref{var inequality for M1}) with respect to $\lambda$.
This gives
\begin{eqnarray*}
2 (1 - \lambda) \> M_{\rm phys}^2 \> - \frac{\partial V}
{\partial \lambda} \> = \> 0.
\end{eqnarray*}
The derivative can be worked out easily (see Eqs. (\ref{pot},
\ref{e(s,t,u)})) and we obtain the implicit equation for $\lambda$
\begin{equation}
\frac{1}{\lambda} = \> 1 + \frac{g^2}{8 \pi^2} \> \int_0^{\infty}
d\sigma\>
\frac{\sigma^2}{\mu^4(\sigma)} \>
\int_0^1 du \> u \> e \> \left ( m \mu(\sigma),
\frac{\lambda M_{\rm phys} \sigma}{ \mu(\sigma)}, u \right)  \>.
\label{var eq for lambda}
\end{equation}
Similarly, the variation with respect to $A(E)$
\begin{eqnarray}
\frac{\delta}{\delta A(E)} \> \left ( \Omega \> + \> V \> \right )
\> = \> 0 \nonumber
\end{eqnarray}
gives
\begin{eqnarray}
A(E) \> = \> 1 + \frac{g^2}{4 \pi^2} \frac{1}{E^2}
\int_0^{\infty} d\sigma \>
\frac{\sin^2 (E \sigma /2)}{\mu^4(\sigma)} \> \int_0^1 &du& \left [
1 + \frac{m^2}{2}
\mu^2(\sigma) \frac{1-u}{u} -\frac{\lambda^2 M^2_{\rm phys} \sigma^2}
{2 \mu^2(\sigma)} u \right ] \nonumber \\
&\cdot& e \> \left ( m \mu(\sigma),
\frac{\lambda M_{\rm phys} \sigma}{ \mu(\sigma)}, u \right) \>,
\label{var eq for A(E)}
\end{eqnarray}
where Eq. (\ref{var eq for amu2})
has been used to evaluate
$\delta \mu^2(\sigma) / \delta A(E) $.

Let us discuss some of the aspects of the coupled variational
equations (\ref{var eq for amu2}) - (\ref{var eq for A(E)}). We
first note that we may read off the retardation function, as
defined in Eq.~(\ref{A(E)}), from the
profile function (\ref{var eq for A(E)}); it is given by
\begin{equation}
f_{\rm var} (\sigma) = \frac{g^2}{32 \pi^2} \>
\frac{1}{\mu^4(\sigma)} \> \int_0^1 du \left [ 1 + \frac{m^2}{2}
\mu^2(\sigma) \frac{1-u}{u} -\frac{\lambda^2 M^2_{\rm phys} \sigma^2}
{2 \mu^2(\sigma)} u \right ] \> e \> \left ( m \mu(\sigma),
\frac{\lambda M_{\rm phys} \sigma}{ \mu(\sigma)}, u \right) \> .
\label{var retardation function}
\end{equation}
Obviously it has the same $1/\sigma^2$-behaviour for small
relative times as the `improved' parametrization
(\ref{improved retard func}). Furthermore, it should be noted that
{\em no renormalization} is needed : all integrals converge for
$\sigma \to 0$. In addition, the variational equations are also
well behaved in the limit $ m \to 0$.
 From Eq. (\ref{var eq for lambda}) we observe that
\begin{equation}
 0 \> < \> \lambda \le \> 1
\end{equation}
always, which allows interpretation of $\lambda$ as a kind of average
``velocity'' (see Eq. (\ref{def lambda})) in the proper time. From
Eq. (\ref{var eq for A(E)}) we find that asymptotically
\begin{eqnarray}
A_{\rm var}(E) \>  \buildrel E \to \infty \over \longrightarrow
\> &1&
+ \> \> \frac{g^2}{4 \pi^2} \frac{1}{E^2} \int_0^{\infty} d\sigma \>
\frac{\sin^2(E \sigma/2)}{\sigma^2} \> + ... \nonumber \\
= \> &1& + \> \>
\frac{g^2}{16 \pi} \frac{1}{E} \> + \> ...
\label{var A(E) for large E}
\end{eqnarray}
which is consistent with Eq. (\ref{improved A(E) for large E}).
Note that while $V$ needs renormalization, $\Omega$ does not because
the $E$-integral in Eq. (\ref{Omega})
is still convergent with the asymptotic behaviour
(\ref{var A(E) for large E}).

\subsection{Approximate solution of the variational equations}
\label{sec: approx solution}

\noindent
Although we will present numerical solutions of the above
variational equations in the following paper, it is very useful
to first attempt to derive some
approximate {\it analytical} results. Because the ratio of the
pion mass compared to the nucleon mass  is small
( $m^2/M^2_{\rm phys} \simeq 0.02 $), a natural approximation to make
is to set the pion mass to zero.  This is a meaningful
thing to do because, as we have already noted,
the variational equations  (\ref{var eq for amu2}) -
(\ref{var eq for A(E)}) are both ultraviolet- and infrared-safe.
For $ m = 0$, the equation for $\lambda$ becomes
\begin{equation}
\frac{1}{\lambda} = \> 1 + \frac{g^2}{2 \pi^2}
\frac{1} {M^4_{\rm phys}
\lambda^4} \> \int_0^{\infty} d\sigma\> \frac{1}{\sigma^2} \>
\left [ 1 - (1 +
\gamma(\sigma) ) \> e^{-\gamma(\sigma)} \right ] \>,
\label{m=0 var eq for lambda}
\end{equation}
and the corresponding equation for the profile function is
\begin{equation}
A(E) = \> 1 \> + \> \frac{g^2}{4 \pi^2} \frac{1}{E^2} \>
\int_0^{\infty} d\sigma\> \frac{\sin^2(E \sigma /2) }{\mu^4(\sigma)}
\> e^{-\gamma(\sigma)}
\label{m=0 var eq for A(E)}
\end{equation}
where
\begin{equation}
\gamma(\sigma) =
\frac{ \lambda^2 M^2_{\rm phys} \sigma^2}{2 \mu^2(\sigma)} \> .
\label{gamma(sigma)}
\end{equation}
Furthermore, as seen in Eqs. (\ref{amu2(sigma small)})
and (\ref{amu2(sigma large)}), the pseudotime $\mu^2(\sigma)$ is
proportional to
$\sigma$ both in the small- and large-$\sigma$  limit.  Let us
for the moment assume, in order
to be able to do the remaining integrals in
Eqs. (\ref{m=0 var eq for lambda})
and (\ref{m=0 var eq for A(E)}),
that the pseudotime is in fact always
proportional to the relative proper time
\begin{equation}
\mu^2(\sigma) \> \approx \> r \> \sigma
\label{amu2 approx}
\end{equation}
with $ r \le 1$. This approximation will be a good one if either
the region of small or large $\sigma$ dominates the integrals.
One can now evaluate all the integrals.
Defining the dimensionless coupling constant
\begin{equation}
\alpha \> = \> \frac{g^2}{4 \pi } \> \frac{1}{M_{\rm phys}^2} \> ,
\label{def alpha}
\end{equation}
the variational equation for $\lambda$
( Eq. \ref{m=0 var eq for lambda}) becomes
\begin{equation}
\frac{1}{\lambda} \> \approx \> 1 \> + \>
\frac{\alpha}{\pi r \lambda^2} \> ,
\label{lambda approx}
\end{equation}
while the variational equation for the profile function yields
\begin{equation}
A(E) \approx \> 1 + { \alpha M^2_{\rm phys}\over 2 E \pi r^2} \>
\left [ \> \arctan {2 r E \over \lambda^2 M^2_{\rm phys}} - \>
{\lambda^2 M^2_{\rm phys}\over 4 E r} \> \ln \left ( 1 +
({2 r E \over \lambda^2 M^2_{\rm phys}})^2
\right ) \> \right ] \> .
\label{A(E) approx}
\end{equation}
In particular
\begin{equation}
A(0) \> \approx \> 1 \> + \> \frac{\alpha}{2 \pi} \>
\frac{1}{\lambda^2 r} = {1 \over 2}(1 + {1 \over \lambda}) \> .
\label{A(0) approx}
\end{equation}
This is precisely the form of the profile function
(\ref{improved A(E)})
obtained with the `improved' retardation function in
Section~\ref{sec: improved}, if we identify
\begin{equation}
C' = {\alpha M_{\rm phys}^2 \over 8 \pi r^2 } \hspace{15mm}
{\rm and}\hspace{15mm}
w = {\lambda^2 M_{\rm phys}^2 \over 2 r}\;\;\;.
\label{w from approx solution}
\end{equation}
Solving Eq. (\ref{lambda approx}) for $\lambda$, one obtains
\begin{equation}
\lambda \> \approx  \> \frac{1}{2} \left [ \> 1 \> \pm \>
\sqrt{1 - \frac{4 \alpha}{\pi r}} \> \right ] \>.
\label{lambda}
\end{equation}
This equation has some rather remarkable properties.  First of all,
it has no real solutions for $\alpha$ larger than
\begin{equation}
\alpha_c \> = \> \frac{\pi}{4} \> r\;\;\;.
\label{alpha crit}
\end{equation}
Below this branchpoint it has two solutions, one approaching
$\lambda = 1$ as the
coupling $\alpha$ goes to zero, while the other one approaches
$\lambda = 0$.
The first of these limits corresponds to the perturbative limit
(see Section ~\ref{sec: on mass shell}), while $\lambda = 0$ seems
 unphysical
(see Eqs. (\ref{var inequality for Mphys}) or
(\ref{var inequality for M1})).

If one argues that mostly small $\sigma$-values
matter in the respective integrals, i.e. $ r \approx 1$ then
\begin{equation}
\alpha_c \> \approx \> \frac{\pi}{4} \> = \> 0.785
\label{num value for alpha crit}
\end{equation}
For $ \alpha > \alpha_c $ only {\em complex} solutions are possible.
This is a sign of the instability of the model and will be studied
in more detail in the following paper.

\section{Discussion and Summary}

\noindent
In this work we have introduced a variational approach to relativistic
quantum field theory which is closely
modelled on the very successful treatment of the polaron in condensed
 matter physics.  The final aim is
to do this for a realistic theory such as QED or a meson-nucleon
theory.
However, there are considerable problems
in going from the non-relativistic polaron problem to a field theory.
So, in order
not to be confronted with all complications
at once, we have chosen to start with a toy theory (the Wick-Cutkosky
 model) which is not a gauge
theory and where spin and isospin degrees of freedom are neglected,
but where the coupling is of a similar Yukawa form as for the more
physically relevant theories mentioned above.
  This theory not only has the advantage of relative
simplicity, but it also turned out that the action is actually
extremely similar to the polaron action so that
one might expect to have similar success by using the same variational
treatment as was introduced by Feynman in the polaron problem.

Following this idea we have integrated out the light mesons and
represented the heavy particles
degrees of freedom by trajectories parametrized by the proper time.
This step necessarily required neglect of heavy particle pair
production, i.e. the quenched approximation. The resulting non-local
effective action $ S_{\rm eff} $ was then
approximated variationally by a retarded quadratic action $ S_t $
whose parameters (the ``profile function'' $ A(E)$ and an average
``velocity'' $ \lambda $)
have to be determined on the pole of the two-point function. Apart
from technical differences the Wick-Cutkosky model here again turned
out to be very similar to the polaron problem. We have introduced two
different ways of averaging over the exact action (``coordinate
averaging and ``momentum averaging'') which gave identical results
on the pole of the two-point function. In contrast to methods which
optimize perturbation theory \cite{SS,SSS} ours is a truly
variational approach and, as shown in the case of
``coordinate averaging'', even a minimum principle.

However,
the model to which we applied our method
clearly also has some disadvantages. One of the technical
differences to the polaron problem is the need of renormalization
in a relativistic field theory. In this respect
the Wick-Cutkosky model is  too simplistic: only a mass
renormalization
is needed in the quenched approximation (i.e. the model is
superrenormalizable) which certainly is not enough
for dealing with the (non-perturbative)
renormalization of realistic theories. Of more immediate concern,
however, is the fact that, unrelated to the variational approach as
such, the model is unstable. This is of
course not a feature of the more realistic problems which one is
interested in the first place. Luckily, we have been
able to ignore this instability in so far as that, at least in
the variational approach presented here, it only starts to manifest
itself for couplings larger than some critical coupling.

Nevertheless, the instability prevents us from comparing the
results of the variational calculation to a strong coupling
limit of the theory.  This is rather unfortunate, as for the
polaron the success of the approach could be
gauged by the excellent agreement of the variational treatment with
{\it both} the strong and weak coupling limits.
Here we can only compare with the latter, a comparison with the
strong coupling limit will have
to wait until the method is applied to a theory where this limit
exists in the first place.  Actually, although
a stable model would have been more welcome, the instability does
allow us, through the use of this non-perturbative method, to explore
the behaviour of the theory around the critical coupling, something
which one could not do in perturbation theory.

As was the case for the polaron, the variational calculation
contains within it first order perturbation theory, as we have seen
by way of example for the self-energy of
the heavy particle. Importantly this is true for {\it any}
value of the variational parameters so that agreement
with the first order perturbative calculation is assured.  In the
language of perturbation theory, variation of the parameters
then allows one to effectively sum up parts of higher diagrams up to
all orders. In principle, the variational approach may be improved
systematically by going beyond the leading order of the cumulant
expansion which is used in Feynman's variational principle. In the
polaron case this leads to results for the ground state energy and
the effective mass \cite{LuRo} which nearly match the exact
Monte-Carlo calculations \cite{AlRo}.
In practise, however, higher order corrections become
increasingly difficult to calculate and so the
usefulness of the approach depends on how closely the leading
orders reflect reality.  In particular,
the accuracy of the zeroth order results (using the first order
variational parameters) is of interest. We will show in a subsequent
paper that already the zeroth order approximation  gives a quite
reasonable description of meson production and scattering processes
after analytic continuation to Minkowski space.

An important ingredient of the approach advocated here is to apply the
variational principle to the action
expressed in terms of particle coordinates rather than fields, as has
previously been done.  The reason for
doing this is the reduction in the number of degrees of freedom which
this entails.  This is important as
one is restricted to generalized quadratic trial actions
for practical variational calculations.
Furthermore, as we have seen, the
particle action makes extraction of the
connected part of a Green function completely trivial.  On the other
hand, one might consider it to be a disadvantage
that the action in the particle representation is non-local.
Although not crucial, there is a certain loss of intuition associated
with this.  For example, in the formulation in terms of fields one
may extend the concept of the classical potential, and the physical
picture which this entails,
to higher orders in the coupling through the use of the effective
potential.  Even at the classical level, it is
immediately clear by looking at the potential in
Fig.~\ref{fig: contourplot}
 that the Wick-Cutkosky model is unstable.
It is rather difficult to see this in the particle representation of
the action  (\ref{eff action pot'}).  Indeed,
even after approximating the particle action by the trial action
one first had to solve a set of nonlinear coupled equations before
any signs of the instability manifested itself.
Fortunately, we could obtain very good approximative results and
analytical insight for the solution of the variational equations by
setting the meson mass to zero and by replacing the ``pseudotime''
$\mu^2(\sigma)$ by its limit
when the relative proper time $\sigma$ tends to zero. The success of
this rather drastic approximation indicates that to a large extent
the dynamical behaviour of this relativistic system is governed by
short-time processes.
Although no substitute for a numerical solution, these analytical
expressions prove to be rather useful
guides to the general behaviour of the solutions.
Whether the value of the
 critical coupling is only an artefact of our present quadratic
approximation or has some physical meaning
is not fully clear. In support of the latter view it may argued that
the critical coupling corresponds to
the situation where the average heavy particle field is just large
enough to overcome the barrier depicted in Fig.~\ref{inverted double well} .

In conclusion, we think that the variational approach in the form
advocated here looks rather promising at least
for the particular model which we have examined.
Not only has it provided rather simple analytical expressions which
go considerably beyond perturbation theory,
but it also allows for numerical investigations which will be
reported in the following paper. We
therefore believe that it is certainly
worthwhile to apply and extend it to other more realistic cases.

\vspace{2cm}

\noindent
{\bf Acknowledgements}

\noindent
We would like to thank Dina Alexandrou and  Yang Lu for stimulating
discussions concerning many aspects of this work
and Geert Jan van Oldenborgh for a careful reading of the manuscript.

\newpage

\noindent
{\Large\bf Appendix : Regularization}

\renewcommand{\theequation}{A.\arabic{equation}}
\setcounter{equation}{0}

\vspace{0.5cm}

\noindent
Here we perform the regularization of the averaged action
$ < \> S_1 \> >_{S_t} $ by subtracting the term
(\ref{Pauli Villars subtraction}) from the $\sigma$-integrand.
Allowing for an arbitrary subtraction point $\mu_0$ we then have
\begin{eqnarray}
< \> S_1 \> >_{S_t} &=& \> - \frac{g^2}{8 \pi^2} \> \int_0^{\beta}
d\sigma
\int_{\sigma/2}^{\beta - \sigma/2} dT \> \int_0^1 du \> \> \Biggl [
\frac{1}{\mu^2(\sigma,T)} \nonumber \\
&\cdot&  e\left ( m \mu(\sigma,T),\>
\frac{x \sigma}{ \beta \mu(\sigma,T)}\> , \> u \right )
- \frac{1}{\sigma} \> e\left ( \Lambda \sqrt{\sigma}, \mu_0
\sqrt{\sigma}, u \right ) \Biggr ] \> .
\label{S1 regularized}
\end{eqnarray}
We write the quantity in square brackets as
\begin{eqnarray}
& &\frac{1}{\sigma} \left [
e\left ( m \sqrt{\sigma}, \mu_0 \sqrt{\sigma}, u \right )
 \> - \> e\left ( \Lambda \sqrt{\sigma}, \mu_0 \sqrt{\sigma}, u
\right ) \right ] \nonumber \\
&+& \frac{1}{\mu^2(\sigma,T)} \>
 e\left ( m \mu(\sigma,T),\>
\frac{x \sigma}{ \beta \mu(\sigma,T)}\>, \> u \right )
\> - \> \frac{1}{\sigma}
e\left ( m \sqrt{\sigma}, \mu_0 \sqrt{\sigma}, u \right )
\end{eqnarray}
and concentrate on the term in the first line which diverges if the
cut-off $\Lambda$ goes to infinity. The term in the second line
gives rise to the regular part (\ref{S1 regular}) of the averaged
action . For the first term we can perform the $T$-integration
immediately since the integrand does not depend on $T$ . This gives
a factor $\beta - \sigma$. With the explicit form (\ref{e(s,t,u)})
of the function $e(s,t,u)$ we then have to evaluate
\begin{equation}
< S_1 >^{\rm div} \> \equiv \> - \frac{g^2}{8 \pi^2} \> \int_0^1 du
\>  \int_0^{\beta} d\sigma \> \frac{\beta - \sigma}{\sigma} \>
\left [ \> e^{ - z_{m,\mu_0}(u) \> \sigma }
\> -\> e^{ - z_{\Lambda,\mu_0}(u) \> \sigma }  \> \right ]
\label{divergent part}
\end{equation}
where
\begin{equation}
z_{m,\mu_0}(u) \> = \> \frac{m^2}{2} \frac{1-u}{u} \> + \>
\frac{\mu_0^2}{2} u \> .
\label{def z(u)}
\end{equation}
The $\sigma$ - integral can be done in terms of the exponential
integral \cite{Handbook}
\begin{equation}
E_1(z) = \int_1^{\infty} dt \> \frac{1}{t} \> e^{-z t} \>.
\label{E1(z)}
\end{equation}
For $ z \to 0 $ this function behaves like
\begin{equation}
E_1(z) \longrightarrow \> - \>\gamma - \ln z - {\cal O}(z)
\label{E1 small z}
\end{equation}
where $\gamma = 0.577215...$ is Euler's number  and for
$z \to \infty$ like
\begin{equation}
E_1(z) \longrightarrow \> \frac{e^{-z}}{z} \> \left [ \> 1 +
{\cal O}\left (\frac{1}{z} \right ) \> \right ] \> .
\label{E1 large z}
\end{equation}
We easily find
\begin{eqnarray}
< S_1 >^{\rm div} &=& - \frac{g^2}{8 \pi^2} \> \int_0^1 du \>
\Biggl \{ \>  \beta \left [ \>
\ln \frac{z_{\Lambda,\mu_0}(u)}{z_{m,\mu_0}(u)} \> + \>
E_1\left (z_{\Lambda,\mu_0}(u) \beta \> \right ) \> - \>
E_1\left (z_{m,\mu_0}(u) \beta \> \right )  \>  \right ] \nonumber \\
&-& \frac{1}{z_{m,\mu_0}(u)} \left [\>  1 - \>
e^{ - z_{m,\mu_0}(u) \beta} \> \right ] \> + \>
\frac{1}{z_{\Lambda,\mu_0}(u)} \left [\>  1 - \>
e^{ - z_{\Lambda,\mu_0}(u) \beta } \> \right ]  \> \Biggr \} \> .
\end{eqnarray}
In the limit where the cut-off mass $\Lambda$ goes to infinity
 this becomes simpler due to Eq. (\ref{E1 large z})
\begin{eqnarray}
< S_1 >^{\rm div} &=& - \frac{g^2}{8 \pi^2} \> \int_0^1 du \> \Biggl
\{ \>  \beta \left [ \> \ln \frac{\Lambda^2}{m^2}   \>  - \>
\ln \left ( 1 + \frac{\mu_0^2}{m^2} \frac{u^2}{1-u} \right ) \> - \>
E_1\left (z_{m,\mu_0}(u) \beta \right )  \> \right ] \nonumber \\
&-& \frac{1}{z_{m,\mu_0}(u)} \left [\>  1 -
e^{ - z_{m,\mu_0}(u) \beta }\> \right ] \Biggr \} \nonumber \\
&\equiv&
- \frac{g^2}{8 \pi^2} \> \beta \ln \frac{\Lambda^2}{m^2} \> + \>
< \> S_1\> >^{\rm fin} \> .
\label{S1 finite}
\end{eqnarray}
The above expression for the finite part simplify considerably for
$\mu_0 = 0$ and/or $\beta \to \infty$. This is what we employ in the
main text. Note that for $m = 0$ we would need $\mu_0 \neq 0$ .

\end{document}